\documentclass[iop,apj]{emulateapj}
\usepackage{color}
\definecolor{darkgreen}{rgb}{0,0.5,0}
\definecolor{magentadye}{rgb}{0.79,0.08,0.48}

\usepackage{graphicx,natbib,amsmath,amssymb}
\usepackage{ulem}
\bibpunct[, ]{(}{)}{;}{a}{}{,}
\newcommand{\simgt}{\hbox{\rlap{\raise 0.425ex\hbox{$>$}}\lower 0.65ex\hbox{$\sim$}}}
\newcommand{\simlt}{\hbox{\rlap{\raise 0.425ex\hbox{$<$}}\lower 0.65ex\hbox{$\sim$}}}

\newbox\grsign \setbox\grsign=\hbox{$>$} \newdimen\grdimen \grdimen=\ht\grsign
\newbox\simlessbox \newbox\simgreatbox
\setbox\simgreatbox=\hbox{\raise.5ex\hbox{$>$}\llap
     {\lower.5ex\hbox{$\sim$}}}\ht1=\grdimen\dp1=0pt
\setbox\simlessbox=\hbox{\raise.5ex\hbox{$<$}\llap 
     {\lower.5ex\hbox{$\sim$}}}\ht2=\grdimen\dp2=0pt
\def\simgt{\mathrel{\copy\simgreatbox}}
\def\simlt{\mathrel{\copy\simlessbox}}

\newcommand{\hMpc}{{\ifmmode{h^{-1}{\rm Mpc}}\else{$h^{-1}$Mpc }\fi}}
\newcommand{\hkpc}{{\ifmmode{h^{-1}{\rm kpc}}\else{$h^{-1}$kpc }\fi}}
\newcommand{\hMsun}{{\ifmmode{h^{-1}{\rm {M_{\odot}}}}\else{$h^{-1}{\rm{M_{\odot}}}$}\fi}}
\newcommand{\Msun}{{\ifmmode{{\rm {M_{\odot}}}}\else{${\rm{M_{\odot}}}$}\fi}}

\newcommand{\dout}{\bgroup \markoverwith{\rule[0.2ex]{0.1pt}{0.4pt}\rule[0.8ex]{0.1pt}{0.4pt}}\ULon}
\newcommand{\ddd}[1]{}

\defcitealias{2010ApJ...722..851H}{Paper I}
\defcitealias{2010ApJ...722..856W}{Paper II}
\defcitealias{2010ApJ...725..282W}{Paper III}

\submitted{Received 2014 February 17; accepted 2015 August 8}
\journalinfo{The Astrophysical Journal, in press (2015)}

\shorttitle{Non-universality of dark-matter halos}
\shortauthors{Hjorth et al.}

\begin{document}
\title{Non-universality of dark-matter halos: 
cusps, cores, and the central potential}


\author{ 
Jens Hjorth\altaffilmark{1}, 
Liliya L. R. Williams\altaffilmark{2},
Rados{\l}aw Wojtak\altaffilmark{1,3}, and
Michael McLaughlin\altaffilmark{2}
}

\altaffiltext{1}{Dark Cosmology Centre, Niels Bohr Institute, University of 
Copenhagen, Juliane Maries Vej 30, DK--2100 Copenhagen \O, Denmark;
jens@dark-cosmology.dk}
\altaffiltext{2}{School of Physics and Astronomy, University of Minnesota, 
116 Church Street SE, Minneapolis, MN 55455, USA; llrw@astro.umn.edu}
\altaffiltext{3}{Kavli Institute for Particle Astrophysics and Cosmology, 
Stanford University, SLAC National Accelerator Laboratory, Menlo Park, 
CA 94025, USA}


\begin{abstract}
Dark-matter halos grown in cosmological simulations appear to have central 
NFW-like density cusps with mean values of $d\log\rho/d\log r \approx -1$,
and some dispersion, which is generally parametrized by the varying index
$\alpha$ in the Einasto density profile fitting function. Non-universality 
in profile shapes is also seen in observed galaxy clusters and possibly 
dwarf galaxies. Here we show that non-universality, at any given mass scale,
is an intrinsic property of DARKexp, a theoretically derived model for 
collisionless self-gravitating systems. We demonstrate that DARKexp---which 
has only one shape parameter, $\phi_0$---fits the dispersion in profile shapes 
of massive simulated halos as well as observed clusters very well. DARKexp 
also allows for cored dark-matter profiles, such as those found for dwarf 
spheroidal galaxies. We provide approximate analytical relations between 
DARKexp $\phi_0$, Einasto $\alpha$, or the central logarithmic slope in the 
Dehnen--Tremaine analytical $\gamma$-models. The range in halo parameters 
reflects a substantial variation in the binding energies per unit mass of 
dark-matter halos.
\end{abstract}

%\baselineskip8pt
%\tableofcontents
%\baselineskip16pt


%% subject header.  Objects should be in the appropriate "individual"
%% headers (e.g. quasars: individual, stars: individual, etc.) with the
%% additional provision that the total number of headers, including each
%% individual object, not exceed six.  The \objectname{} macro, and its
%% alias \object{}, is used to mark each object.  The macro takes the object
%% name as its primary argument.  This name will appear in the paper
%% and serve as the link's anchor in the electronic edition if the name
%% is recognized by the data centers.  The macro also takes an optional
%% argument in parentheses in cases where the data center identification
%% differs from what is to be printed in the paper.



\section{Introduction}

The basic shape of N-body simulated cold dark matter halos is now well 
established. The NFW density profile \citep{1997ApJ...490..493N} is a 
remarkably good representation
\citep[e.g.,][]{2004ApJ...607..125T,2004MNRAS.353..624D,2005Natur.433..389D,2005MNRAS.364..665D,2010MNRAS.402...21N,2011ASL.....4..297D,2011MNRAS.415.3895L}, 
especially given its simplicity: it has no shape parameters. However, a closer 
examination of the profile shapes of relaxed halos shows that NFW does not 
fully capture the density run. Most importantly, while the NFW profile has 
a central density cusp, with logarithmic slope $d\log \rho(r)/d\log r = -1$, 
recent high resolution simulations have shown that the central cusp tends to 
become somewhat shallower at smaller and smaller radii 
\citep[e.g.,][]{2010MNRAS.402...21N}, prompting the use of an Einasto 
$\alpha$-model \citep{2004MNRAS.349.1039N,1965TrAlm...5...87E}. The Einasto 
profile fits halos better at most radii, not just in the center, suggesting 
that the additional parameter $\alpha$ is a real characteristic of dark 
matter halos \citep{2006AJ....132.2685M,2014arXiv1411.4001K}.

\citet{2013MNRAS.432.1103L} find that the mean value of $\alpha$ ranges from 
about 0.17 for $M_{200}=10^{11}$ $h^{-1}$ M$_\odot$ halos to $0.21$ at 
$M_{200}=10^{15}$ $h^{-1}$ M$_\odot$, which is consistent with 
\citet{2008MNRAS.387..536G} who find a relation between Einasto's $\alpha$ 
and the ratio of the linear density threshold for collapse to the rms linear 
density fluctuation at a given mass scale. In addition to the mass-dependent 
$\alpha$, there is significant scatter of around $0.05$ in $\alpha$. Moreover, 
halos with $\alpha$ in the range $0.05$ to $0.45$ are also realized in 
simulations \citep{2013MNRAS.432.1103L,2014arXiv1411.4001K}. These works show 
that the shape of the density profile, including the central logarithmic 
slope, is tied to the initial conditions and evolutionary history of the 
halos in simulations. \citet{2013MNRAS.432.1103L} suggest that the mass 
profiles of halos reflect their accretion history.

The spread in $\alpha$ index is observed not just for main halos, but also
subhalos. \citet{2008MNRAS.391.1685S} find that the profiles of subhalos do 
not converge to a central power law but exhibit curving shapes well described 
by the Einasto form with $\alpha$ in the range $0.15$ to $0.28$. This is 
consistent with the findings of \citet{2013MNRAS.428.1696V}, who note that 
simulated satellite dark matter halos around galaxies are better fit with 
Einasto profiles of index $\alpha=0.2-0.5$. \citet{2013MNRAS.431.1220D} find 
a correlation between $\alpha^{-1}$ (ranging from 0.1 to 1) and satellite 
dark matter halo mass, i.e., with higher masses corresponding to smaller 
values of $\alpha$, contrary to the trend found for more massive halos. The 
diversity in the density profiles of these satellites is attributed to tidal 
stripping, which acts mainly in the outer parts of the subhaloes. 

The conclusion is that simulated dark matter halos do not form a strictly 
self-similar family of systems as was originally suggested, but display 
non-universality \citep{2006AJ....132.2685M}.

Neither NFW nor Einasto have any theoretical underpinnings; they are purely 
empirical. In a series of papers 
\citep{2010ApJ...722..851H,2010ApJ...722..856W,2010ApJ...725..282W,2014ApJ...783...13W}
we have proposed a theoretically based model, DARKexp, which has been shown 
to allow good fits to a few simulated halos 
\citep{2010ApJ...725..282W,2014ApJ...783...13W}
and observed galaxy clusters \citep{2013MNRAS.436.2616B,2015arXiv150704385U}.

DARKexp has one shape parameter, and its density profiles are very similar 
to Einasto within a limited range of shape parameters. Taking the broader 
range of achievable values into account, DARKexp suggests that there should 
be a dispersion in the shapes of the density profiles between systems, 
parametrized by the shape parameter. At small radii, but sufficiently
above the resolution limit achieved in cosmological simulations, these density 
profiles differ significantly. Some have nearly flat cores, while others are 
steep, $d\log \rho(r)/d\log r \approx -2$. In other words, unless halos have 
identical normalized central potentials, DARKexp predicts that the density 
profiles should not be universal, and that a logarithmic slope of $-1$ 
at small radii robustly resolved in simulations is not a universal attractor 
of halos with different formation histories. Testing this would be important 
because DARKexp is not a fitting function but is derived from basic physics. 

We summarize the rationale and properties of DARKexp in Section~\ref{darkexp}.
While DARKexp has the advantage of being derived from first principles, the 
resulting density profiles unfortunately cannot be expressed analytically.
We therefore also relate the properties of DARKexp to those of phenomenological 
fitting functions, i.e., generalized Hernquist, Dehnen--Tremaine, and Einasto 
models. We discuss non-universality of simulated and observed massive
dark-matter halos in Section 3 and show that, in cases where exact fits 
are not required, DARKexp, Einasto, or Dehnen--Tremaine profiles provide 
equally good representations, with considerable dispersion in their respective 
shape parameters. We relate observed, simulated and theoretical properties of 
%lower mass halos in Section~\ref{dwarfs} and conclude in Section~\ref{conc}.
lower mass halos in Section~4 and conclude in Section~5

\section{DARK{\lowercase {exp}}
and the central potential--density slope relation}\label{darkexp}

DARKexp is a theoretical model for the differential energy distribution in a 
self-gravitating collisionless system. Specifically, defining 
$N(\epsilon) = dM/d\epsilon$, where $\epsilon$ is the normalized absolute 
particle energy per unit mass, a statistical mechanical analysis leads to
\begin{equation}
N(\epsilon)  \propto \exp(\phi_0-\epsilon)-1
\end{equation}
\citep{2010ApJ...722..851H}. The central dimensionless potential, $\phi_0$, is 
the only free shape parameter of the model. Some of the global properties of 
the model are analytic (see Appendix A), while the phase space distribution 
function and derived properties, such as the density profile, must be obtained 
numerically using an iterative procedure \citep{2010ApJ...725..282W}. Apart 
from the assumption of isotropy (which leads to spherical structures), there 
are no additional assumptions involved in this computation. The scale 
parameters of the corresponding density profile follow from the resulting 
profiles themselves. DARKexp is a maximum entropy theory and as such describes 
only the final, equilibrium states of systems, and not the evolutionary paths 
that brought them to the final state. The density profiles apply to isolated 
systems, i.e., there is no cosmological context and so no concept of a virial 
radius or formation redshift.

While the asymptotic logarithmic density slope is $-1$ for $r\to 0$
\citep{2010ApJ...722..851H}, in reality the central logarithmic slope 
robustly resolved in simulations depends on $\phi_0$: $\phi_0\lesssim 4$ 
systems have inner logarithmic slopes shallower than $-1$, while 
$\phi_0\gtrsim 5$ have inner logarithmic slopes between $-1$ and $-2$ 
\citep{2010ApJ...722..856W}. DARKexp was not designed to fit the properties 
of dark-matter structures. Nevertheless, the energy distributions resulting 
from this first principles derivation are excellent representations of 
numerical simulations and the density profiles are reminiscent of Einasto 
$\alpha$ profiles for values of $\phi_0$ around 4 \citep{2010ApJ...725..282W}. 
Note that in the \citet{2014ApJ...783...13W} formalism, the DARKexp density 
run is independent of any velocity anisotropy. DARKexp also provides very good 
fits to observed galaxy clusters 
\citep{2013MNRAS.436.2616B,2015arXiv150704385U}, and globular clusters 
\citep{2012MNRAS.423.3589W}. 

We here compare DARKexp to phenomenological fitting functions, notably the 
generalized Hernquist, the special case of the Dehnen--Tremaine $\gamma-$model, 
and the Einasto profile (see Appendix B for details).\footnote{For convenience 
we are making numerically generated models for $\phi_0 = 1$--10 available for 
download at the DARKexp website http://www.dark-cosmology.dk/DARKexp/.}
We provide the correspondence between the shape parameters of these analytic 
fitting functions and the dimensionless parameter of DARKexp, $\phi_0$. These 
results can be used in any situation where fitting with DARKexp is required.

By construction, the DARKexp energy distribution is discontinuous at the escape 
energy, and as a consequence the density falls off as $r^{-4}$ at large radii 
\citep{1987IAUS..127..511J,1991MNRAS.253..703H}. When searching for
phenomenological or analytical profiles that may be useful approximations to 
DARKexp, we recall that the generalized \citet{1990ApJ...356..359H} model 
has the same outer logarithmic slope of $-4$ as DARKexp, and also allows for 
a range of central logarithmic slopes,
\begin{equation}\label{hernmod}
\rho \propto r^{-\gamma}(1+r^\beta)^{(\gamma-4)/\beta} ,
\end{equation}
where $\beta$ governs the sharpness of the transition between the asymptotic 
inner logarithmic slope of $-\gamma$ and the outer logarithmic slope of $-4$. 
For $\beta=1$ the resulting $\gamma$-models,
\begin{equation}\label{gamod}
\rho \propto r^{-\gamma}(1+r)^{\gamma-4},
\end{equation}
are largely analytical \citep{1993MNRAS.265..250D,1994AJ....107..634T}.

Another phenomenological model, which provides an excellent fit to numerical 
simulations, is the Einasto model \citep{2004MNRAS.349.1039N}; it is 
controlled by the shape parameter $\alpha$,
\begin{equation}\label{einmod}
\rho \propto \exp\left [-\frac{2}{\alpha}(r^\alpha-1)\right].
%slope=-2*xx^alpha
\end{equation}
As shown by \citet{2010ApJ...722..856W} and \citet{2010ApJ...725..282W}, 
DARKexp halos have a range of logarithmic slopes at small radii robustly 
resolved in simulations depending on $\phi_0$; models in a range around 
$\phi_0\approx4$ are well represented by Einasto models with $\alpha$ in the 
range 0.15--0.2. 

To investigate the relations between the shape parameters describing DARKexp 
and phenomenological models we now turn to a detailed comparison of their 
density profiles (see Appendix B). For DARKexp of a given 
$\phi_0$ we find an Einasto model or $\gamma$-model that matches DARKexp 
(i.e., has the same value as DARKexp) in either density or log--log density 
slope. The anchor radius at which the matching is done is chosen to be 
$\log(r_a/r_{-2}) = -1.5$, which is the typical resolution limit in 
cosmological simulations \citep[e.g.,][]{2010MNRAS.402...21N}. We choose to 
anchor the profiles at a fixed radius rather than fitting the (log) density or 
slope as a function of (log) radius, over some range, for visual clarity.

\begin{figure}[h]%[t]
\epsscale{1.0}
\plotone{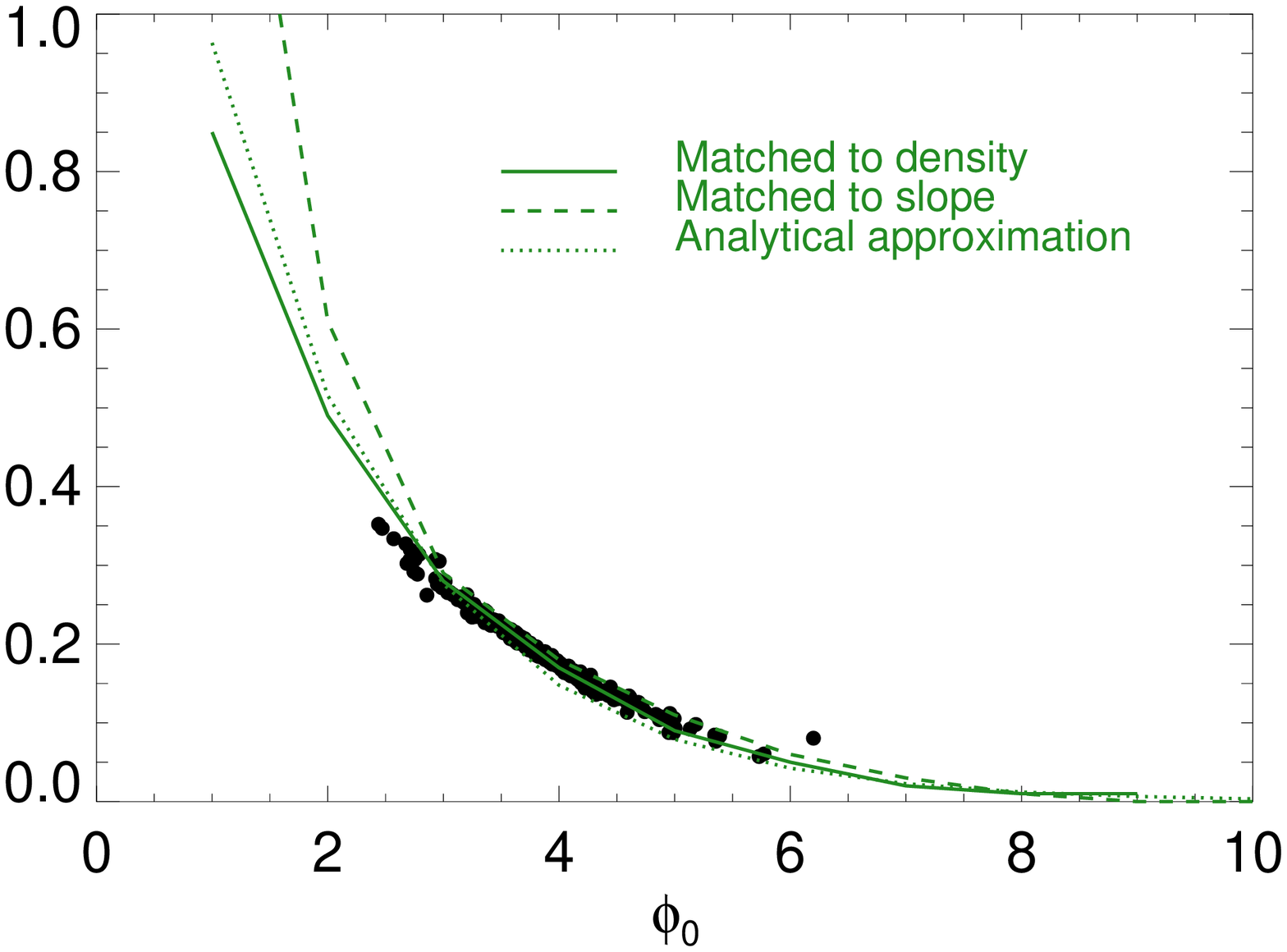}
\plotone{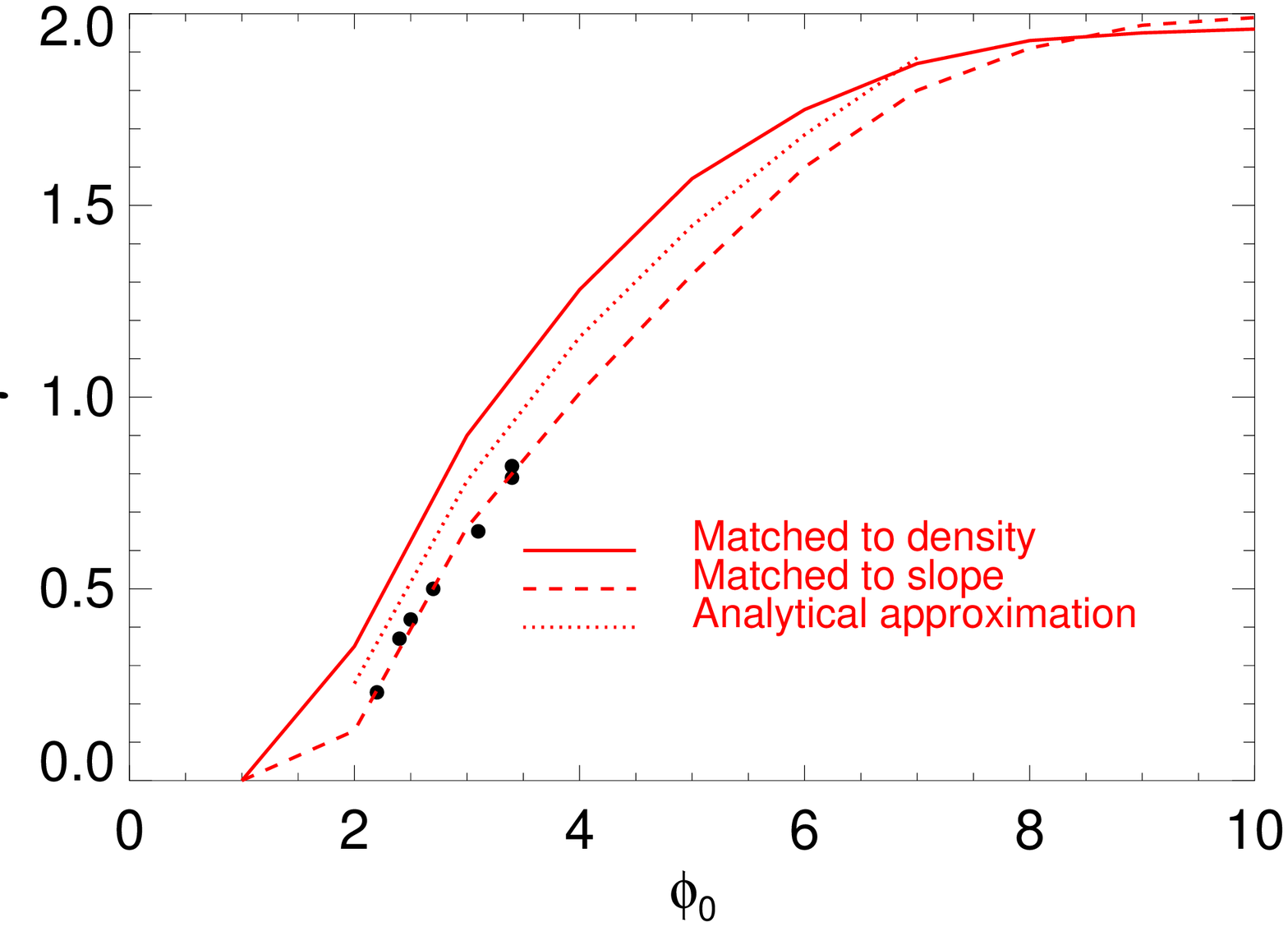}
\caption{$\phi_0$--$\alpha$ relation (top) and $\phi_0$--$\gamma$ relation 
(bottom) for DARKexp. The solid curves are for match in density, the dashed 
curve are for match in logarithmic slope at $\log (r_a/r_{-2}) = -1.5$. The 
dotted curves represent analytical approximation to these relations (Equations 
\ref{gamma-approx} and \ref{alpha-approx}). In the top panel, the black 
symbols represent simulated halos (Section 3.1, Figure~\ref{density-sim}). 
In the bottom panel, the black symbols are clusters of galaxies from 
\citet{2013ApJ...765...25N}, obtained as discussed in the text (Section 3.2).
}
\label{phi0-gamma}
\end{figure}

\begin{figure}[h]%[t]
\epsscale{1.00}
\plotone{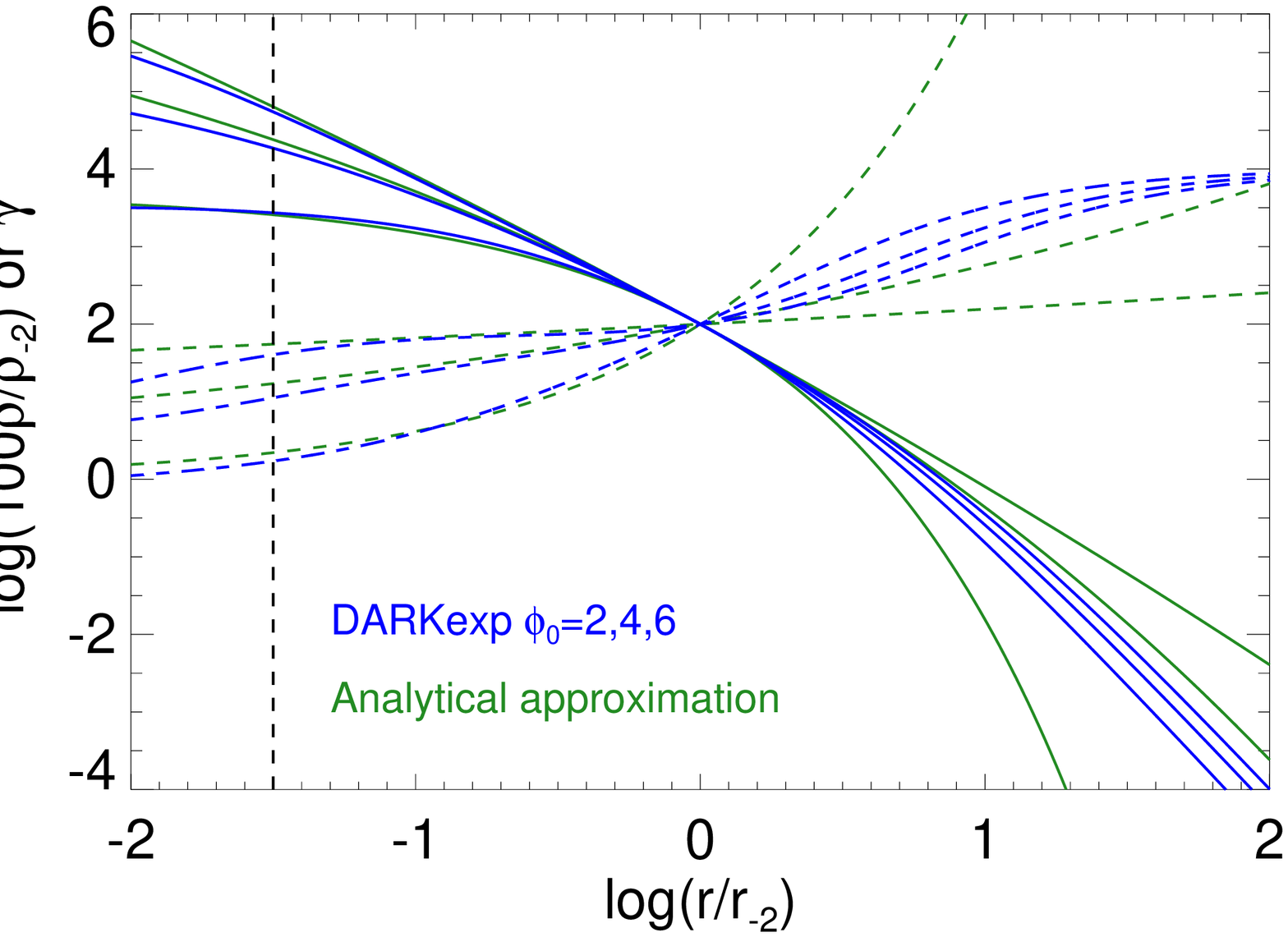}
\plotone{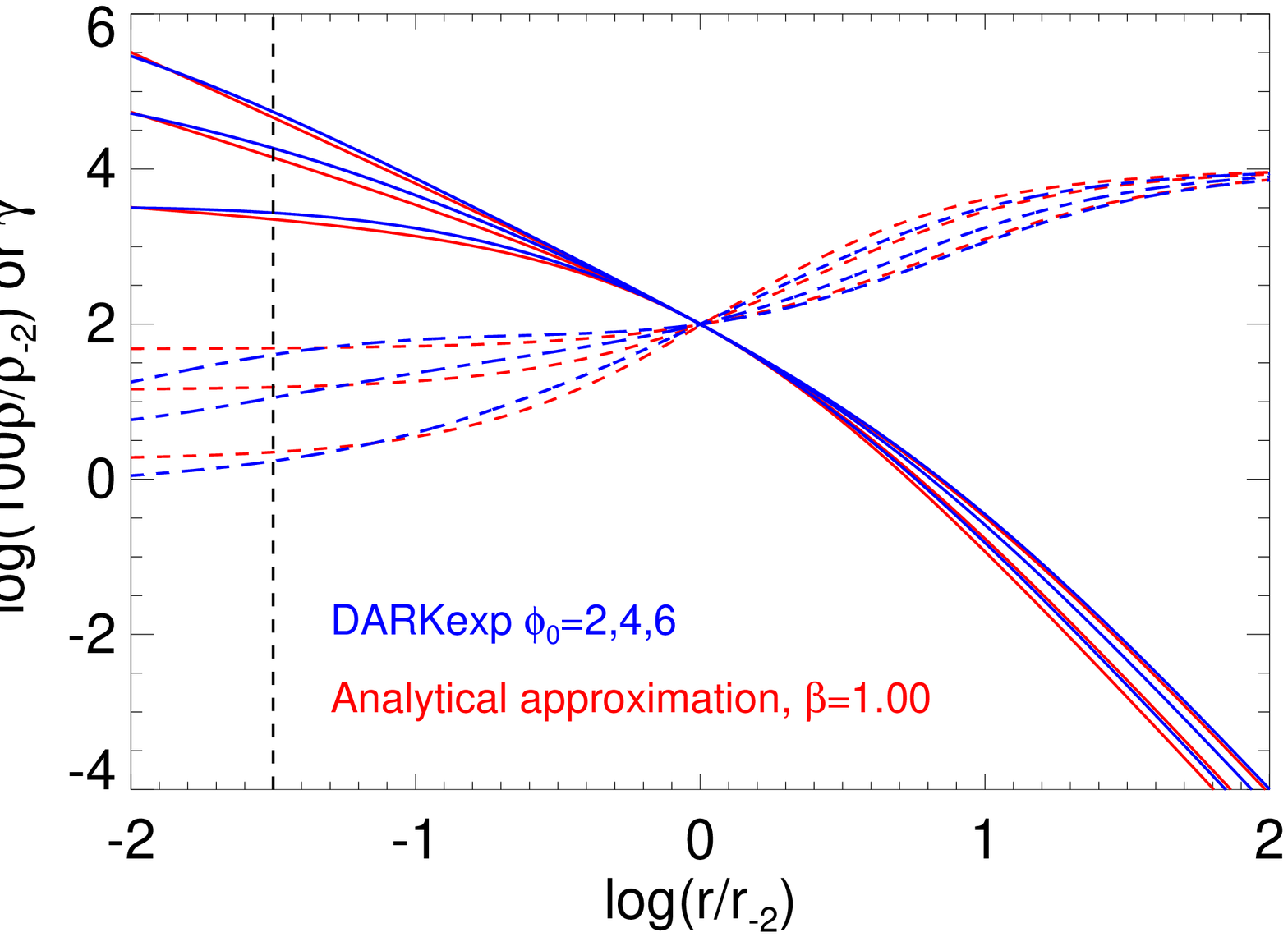}
\caption{
Comparison of DARKexp with Einasto (top) and $\gamma$-models (bottom), using 
the analytical $\phi_0$--$\alpha$ relation (Equation \ref{alpha-approx}, top 
panel) or the analytical $\phi_0$--$\gamma$ relation 
(Equation~\ref{gamma-approx}, bottom panel). DARKexp models with 
$\phi_0=2,4,6$ (from bottom to top at small radii) are shown in blue (solid 
curves: density profile; dashed curves: logarithmic density slope).}
\label{analytical}
\end{figure}

Figure~\ref{phi0-gamma} summarizes the results of the matching between DARKexp 
and phenomenological models. It shows the relations between $\phi_0$ and 
either $\gamma$ or $\alpha$, matched in either density or logarithmic slope. 
We also provide analytical approximations to these relations. For the 
$\gamma$-model
\begin{equation}
\label{gamma-approx}
\gamma\approx 3\log \phi_0 -0.65; \ \ \ \ \ \ \ \ \ \ 1.7\le\phi_0\le 6,
\end{equation}
which suggests $\phi_0\approx 3.5$ for $\gamma=1$. For the Einasto model
\begin{equation}
\label{alpha-approx}
\alpha\approx 1.8\exp(-\phi_0/1.6)
\end{equation}
and $\phi_0\approx 3.8$ for $\alpha=0.17$.

Comparisons between DARKexp and the empirical models, using the above 
analytical relations, are shown in Figure~\ref{analytical}. We will return 
to the central potential--density slope relation in the following sections.

\section{Non-universality of massive halos}
\subsection{Simulated dark matter halos\label{DMsim}}

Until about 5--10 years ago simulation results were consistent with halos 
being universal in shape. Now, there is considerable evidence to the contrary. 
How does the DARKexp expectation of non-universality compare to simulations?
We here analyze recent simulations, show that DARKexp fits the density 
profiles at least as well as Einasto, and quantify the non-universality.

We use cluster halos identified in the Bolshoi simulation\footnote{The 
simulation is publicly available through the MultiDark database 
(http://www.multidark.org). See \citet{2013AN....334..691R} for details of 
the database.}, a dark matter-only simulation run in a $250\hMpc$ box
(WMAP5 cosmological parameters) with $2048^{3}$ particles, which corresponds 
to a mass resolution of $1.35\times10^{8}\hMsun$ \citep[for more details, 
see][]{2011ApJ...740..102K}. For the purpose of our analysis, we select 
well-resolved equilibrium halos with at least $5\times10^{5}$ particles within 
the virial radius $r_{\rm v}$ (defined as the radius of a sphere, whose 
enclosed density is $360$ times larger then the background density). The 
state of equilibrium is assessed using two diagnostics: the virial ratio and 
the relative offset between the minimum of the potential and the halo mass 
center. We find $261$ equilibrium halos defined by the virial ratio less than 
$1.3$ and the mass center offset less than $0.05r_{\rm v}$, which are two 
commonly used relaxation criteria proposed by \citet{2007MNRAS.381.1450N}. 
This selection is done in order to make a meaningful comparison with DARKexp, 
which by definition only applies to equilibrium systems.

\begin{figure*}%[t]
\epsscale{0.9}
\plotone{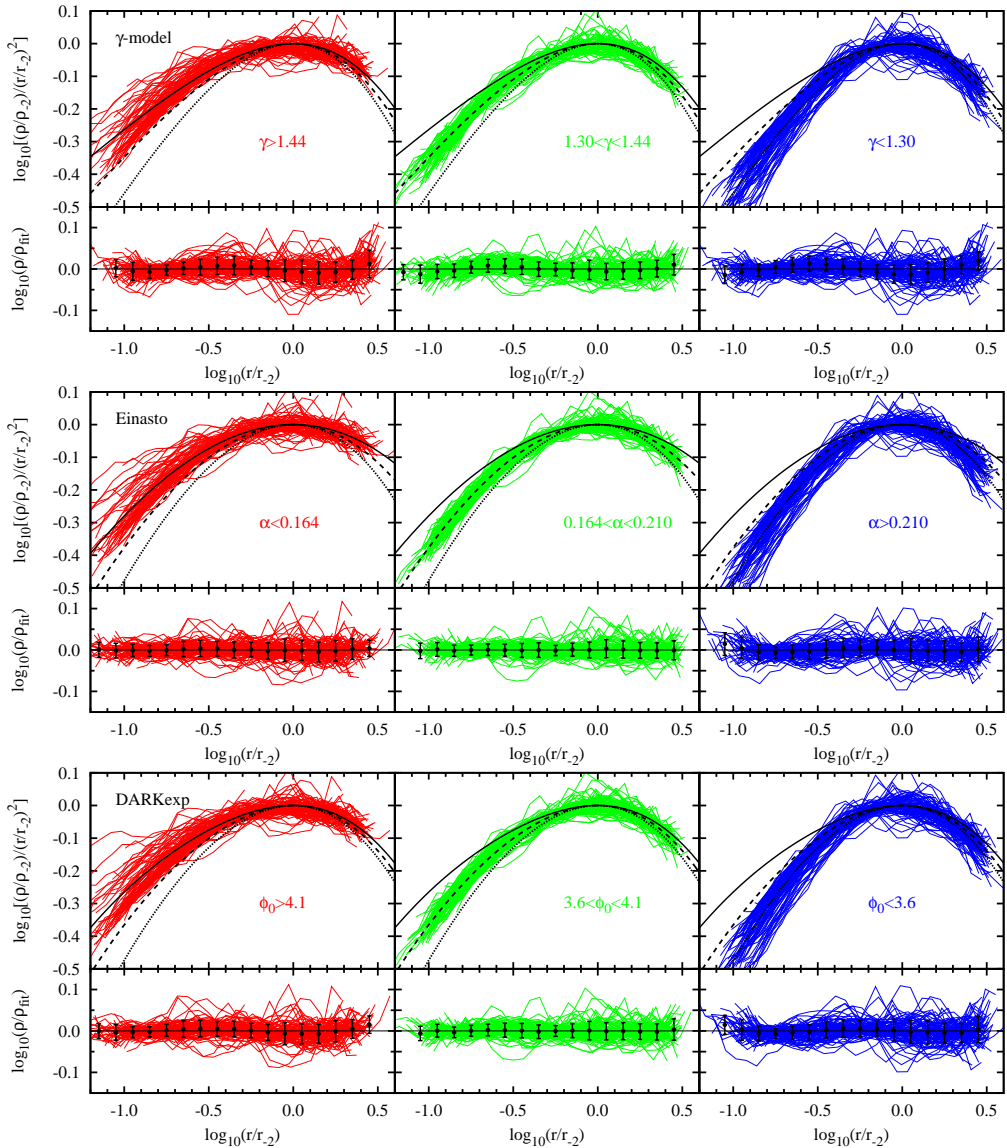}
\caption{
Spherically averaged density profiles of cluster-sized equilibrium dark matter 
halos formed in cosmological simulations. The profiles are split into three 
subsamples (columns) corresponding to three ranges of the best-fit shape 
parameter $\alpha$ of the $\gamma$ profile (top), Einasto profile (middle) or 
the best-fit central potential $\phi_{0}$ of DARKexp (bottom). The black 
curves are the $\gamma$ (top), Einasto (middle), or DARKexp (bottom) profiles 
for the mean $\gamma$, $\alpha$ or $\phi_{0}$ in the corresponding halo sample 
(solid curves: $\gamma > 1.44$, $\alpha < 0.164$, $\phi_0 > 4.1$; 
dashed curves: $1.30 < \gamma < 1.44$, $0.164 < \alpha < 0.210$, 
$3.6 < \phi_0 < 4.1$; dotted curves: $\gamma < 1.30$, $\alpha < 0.210$, 
$\phi_0 < 3.6$). The lower panels in either row show the residuals from the 
best fitting profiles. DARKexp provides as good fits to the simulated halos as 
the Einasto profile and accounts for the range in properties revealed here.
}
\label{density-sim}
\end{figure*}

We calculate the density profiles in $30$ spherical shells equally spaced in 
a logarithmic scale of radius spanning the range from $r_{\rm min}$ to 
$r_{\rm v}$, where $r_{\rm min}=0.015r_{\rm v}$ is the convergence radius 
defined by \citet{2003MNRAS.338...14P} and estimated for the minimum number 
of particles per halo. The halo centers are given by the minimum of the 
potential. DARKexp, Einasto, and Dehnen--Tremaine models are fitted to 
the density profiles of the halos by minimizing the following function
\begin{equation}
\chi=\sum_{i=1}^{n_{\rm shell}}[\ln\rho_{i}-\ln\rho_{\rm model}(r_{i})]^{2},
\end{equation}
where $n_{\rm shell}$ is the number of shells and $r_{i}$ is the mean radius of 
particles inside the $i$-th shell. Following \citet{2011MNRAS.415.3895L}, we 
restrict all fits to the radial range $r_{\rm min}<r<3r_{-2}$, which prevents 
including dynamically less relaxed parts of the halos. Here, $r_{-2}$ is 
defined as the radius at which $d\log \rho(r)/d\log r = -2$, and 
$\rho_{-2}=\rho(r_{-2})$.  We first estimate $r_{-2}$ by fitting an Einasto 
profile in all $30$ shells, i.e., within the radial range 
$r_{\rm min}<r<r_{\rm v}$. This sets the restricted fitting range to be the 
same in both cases. Next, $r_{-2}$, $\rho_{-2}$, and $\gamma$, $\alpha$ or 
$\phi_{0}$ are obtained from a restricted fit to either model. In fitting 
DARKexp, we used density profiles computed on a grid of $\phi_{0}$, ranging 
from 2.0 to 8.0 with an interval of 0.01.

Figure~\ref{density-sim} shows the density profiles of all halos selected for 
the analysis, in three ranges of the shape parameter, i.e., $\alpha$ for the 
Einasto model, $\gamma$ for the Dehnen--Tremaine model, and $\phi_{0}$ for 
DARKexp. The ranges were chosen such that there are equal numbers of halos in 
each of the three bins. The mean value of $\gamma$ is 1.36 with an rms scatter 
of 0.18 while the mean value of $\alpha$ is 0.19 (rms scatter 0.05). For 
$\phi_0$ the mean value is 3.9 (rms scatter 0.6). There is no significant 
correlation in these parameters with the mass over the narrow range of masses 
considered ($10^{14}$ -- $10^{14.8}$ M$_\odot$). The lack of any difference 
between residuals from the best-fitting Einasto, Dehnen--Tremaine, and DARKexp 
profile leads to the conclusion that \textit{these models recover spherically 
averaged density profiles of simulated cosmological halos equally well.} 

The second important conclusion is that, in the narrow mass range considered,
\textit{there is a range of density profile shapes among simulated systems, 
and that the DARKexp family of models captures that spread very well}. 
Figure~\ref{phi0-gamma} (top) illustrates this pictorially; it shows the 
best-fit shape parameters of the Einasto and DARKexp fits to the simulated 
halos; $\phi_0$ and $\alpha$ range between 2.5--6 and 0.05--0.35, respectively 
($\gamma$ ranges from 0.8 to 1.7). The scatter in $\alpha$ is about 0.05.
As expected (Section 2), the data show a tight relation between the central 
potential $\phi_{0}$ and the shape parameter $\alpha$ of the Einasto profile, 
which traces the $\alpha$--$\phi_{0}$ relations obtained from matching Einasto 
and DARKexp density profiles.

\subsection{Observed clusters of galaxies\label{DMobs}}

\citet{2008A&A...489...23L} have emphasized the wide range of central 
logarithmic slopes inferred observationally for different clusters of 
galaxies. DARKexp applies to equilibrium collisionless systems, so in addition 
to dark matter halos it can also be applied to galaxies and clusters that host 
baryons, regardless of the physical processes that led to the formation of the 
final equilibrium systems, provided that the evolution and the present state 
of these systems are not dominated by two-body encounters, and the systems are 
not far from being spherical. We here show that relaxed galaxy clusters 
exhibit non-universality that can be described by the DARKexp model.

\citet{2013ApJ...765...25N} constructed spherically averaged density profiles 
of seven massive relaxed clusters, using central galaxy kinematics on the 
smallest scales, strong lensing on intermediate scales and weak lensing on 
large scales. To separate out dark matter density profiles they subtracted the 
stellar contribution of the central galaxy assuming constant mass-to-light 
ratio for the stellar population. The remaining dark matter profiles were fit 
with cored NFW and generalized NFW (gNFW) forms.

The $\gamma$-model we use here and the gNFW differ in the asymptotic outer 
logarithmic slope ($-4$ vs.\ $-3$), so the differences are confined mostly to 
radii outside $r_{-2}$. At $r\lesssim r_{-2}$ the two models are parametrized 
similarly, therefore one can compare the two models, if one leaves out the 
outermost radii. The \citet{2013ApJ...765...25N} cluster profiles extend about 
2 decades in radius interior to $r_{-2}$, and about 0.75 of a decade exterior 
to $r_{-2}$, so the bulk of the radial range is usable in the comparison.

We took the \citet{2013ApJ...765...25N} inner logarithmic slope for each 
cluster and set it equal to $\gamma$. This supplies the vertical axis 
coordinate in Figure~\ref{phi0-gamma} (bottom). We then fit DARKexp directly to 
\citet{2013ApJ...765...25N} gNFW fits, excluding the outer radii beyond which 
the logarithmic slope is steeper than $-2.7$ (the fits are insensitive to 
the truncation logarithmic slope for values between $-2$ and $-2.8$). This 
gave us $\phi_0$, plotted along the horizontal axis. The seven galaxy clusters 
are plotted as solid points in Figure~\ref{phi0-gamma} (bottom). They fit well 
the $\phi_0$--$\gamma$ relation matched in slope and are consistent with our 
proposed analytical approximation (the dotted red line).

From Figure~\ref{phi0-gamma} we conclude that clusters are fit with a range of 
$\phi_0$'s, suggesting non-universality. Observed relaxed clusters of galaxies 
appear to follow DARKexp; and seem to prefer $\phi_0$ values about 2--3.5, on 
average smaller than cluster-sized dark matter halos in pure dark matter 
simulations, which are fitted with values of $\phi_0$ in the range 2--6.

\section{Dwarf galaxies\label{dwarfs}} 

So far we discussed dispersion in density profiles among massive pure dark 
matter halos. We next consider the differences between pure dark matter halos 
and observed low mass systems.

The central density profiles of observed galaxy clusters appear to be, on 
average, shallower than those of clusters in pure dark matter simulations. 
Likewise, the central logarithmic slopes of observed dwarf galaxies are 
shallower than simulated dark-matter subhalos. In this section we discuss the 
relative energies between the steeper pure dark matter and shallower observed 
structures. We will show that the differences amount to a large fraction of 
the total binding energies of the haloes, demonstrating that the differences
in central logarithmic slopes or $\phi_0$ represent significant differences 
between the halos in terms of global energetics. Moreover, assuming the 
difference between two states or halos is brought about by some dynamical or 
baryonic proces, we compute what it would take to induce a transformation 
between the two, such as the cusp--core transition. Calculating relative 
(fractional) energetics allows us to bypass the uncertainties and unknowns 
associated with the specific physical processes that would bring about the 
transformation.

\subsection{Lack of cusps in dwarf galaxies}

The differences between the NFW and Einasto descriptions of pure dark matter 
halos are small in comparison to the differences between dark matter profiles 
of subhalos/satellite halos and observed dwarf galaxies which may have central 
cores. Most dwarf galaxies appear to exhibit significant central density cores
\citep[][]{1994ApJ...427L...1F,1994Natur.370..629M, 2006ApJ...652..306S,2007ApJ...663..948G,
2009ApJ...704.1274W,2011MNRAS.415L..40B,2011ApJ...742...20W,2012MNRAS.419..184A,2013MNRAS.429L..89A,
2013arXiv1309.2637J,2014ApJ...783....7C}.
%ZZZ Jardel & Gebhardt is published but not in ADS
Whether all dwarfs have a universal profile or not is still unclear because of 
the large uncertainties associated with the measurement of the central 
logarithmic density slopes \citep[e.g.,][]{2014arXiv1406.6079S}; conclusions 
seem to depend on the sample and the methods used. For example, 
\citet{2014ApJ...789...63A} claim that dwarfs are consistent with having a 
universal central logarithmic density slope of $-0.63$, with a scatter of 
0.28, while \citet{2013arXiv1309.2637J} claim that individual dwarf spheroidals 
show considerable scatter around an average logarithmic slope of $-1$.
 
For reviews of this small scale controversy, see \citet{2013arXiv1306.0913W}
and \citet{2014Natur.506..171P}. The suggested solutions to this set of 
problems are either baryonic or particle in nature. Warm dark matter does not 
seem to be the solution: \citet{2012MNRAS.424.1105M,2013MNRAS.428.3715M} and 
\citet{2013MNRAS.430.2346S} find that warm dark matter leads to cusps just 
like cold dark matter does, although with very small central cores which 
however are astrophysically uninteresting and much smaller than what is 
claimed to be present in dwarf satellites. Indeed, \citet{2014MNRAS.439..300L} 
conclude that warm dark matter halos and subhalos have cuspy density 
distributions that are well described by NFW or Einasto profiles 
\citep[see also][]{2012MNRAS.420.2318L}. \citet{2012MNRAS.421.3464P} and 
\citet{2014ApJ...792...99S} argue that baryonic physics 
\citep{2012MNRAS.422.1231G}, via a process similar to violent relaxation, can 
lead to changes in the potential, specifically, a more shallow potential and 
hence a core \citep[see also][]{2012MNRAS.424.1275B,2012ApJ...761...71Z}.
Alternatively, self-interacting dark matter can produce large cores 
\citep{2012MNRAS.423.3740V,2013MNRAS.430...81R,2013MNRAS.431L..20Z,
2014MNRAS.444.3684V}.

\subsection{Cusp--core transition}

Real systems like dwarf spheroidals may have shallower central logarithmic 
slopes and larger central regions of near constant density than pure dark 
matter halos. How does one make the central potential less deep than found in 
numerical simulations? The change is likely brought about by baryonic physics 
\citep{1996MNRAS.283L..72N,2012MNRAS.422.1231G,2015arXiv150202036O} which 
may induce a transformation from a typical $-1$ logarithmic slope to a cored 
halo \citep{2012ApJ...759L..42P}. In our picture this would correspond to a 
change in the central potential, through some energy injection process. 
For DARKexp to be valid, this energy injection would have to be collisionless 
to preserve the assumptions behind the DARKexp derivation. It is interesting 
to note that the baryonic feedback process proposed by 
\citet{2012MNRAS.421.3464P} is not only collisionless but also dissipationless.

We here illustrate the cusp-core transition as a transformation between two 
DARKexp halos.  We can use DARKexp as a starting point for modeling baryonic 
effects if they are mostly non-collisional, in which case DARKexp is also a 
good descriptor of the final state. Note, however, that we do not intend to 
model the details of the transition in terms of detailed physics 
\citep[e.g.,][]{2015arXiv150202036O} such as the mass accretion history 
or star-formation history of the system, the supernova energy deposition 
efficiency, or the stellar initial mass function.

\begin{figure}[h]%[t]
\epsscale{1.0}
\plotone{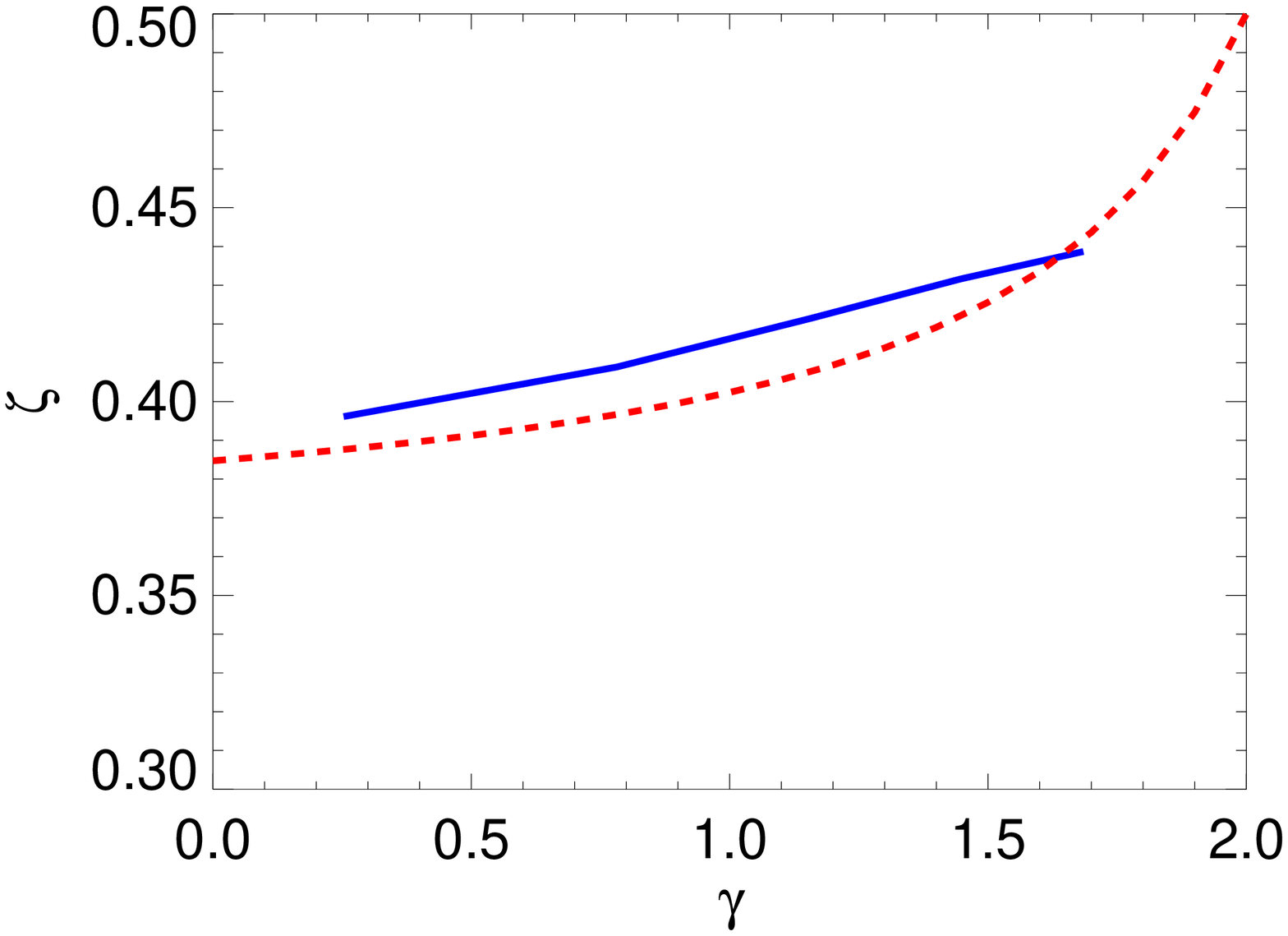}
\plotone{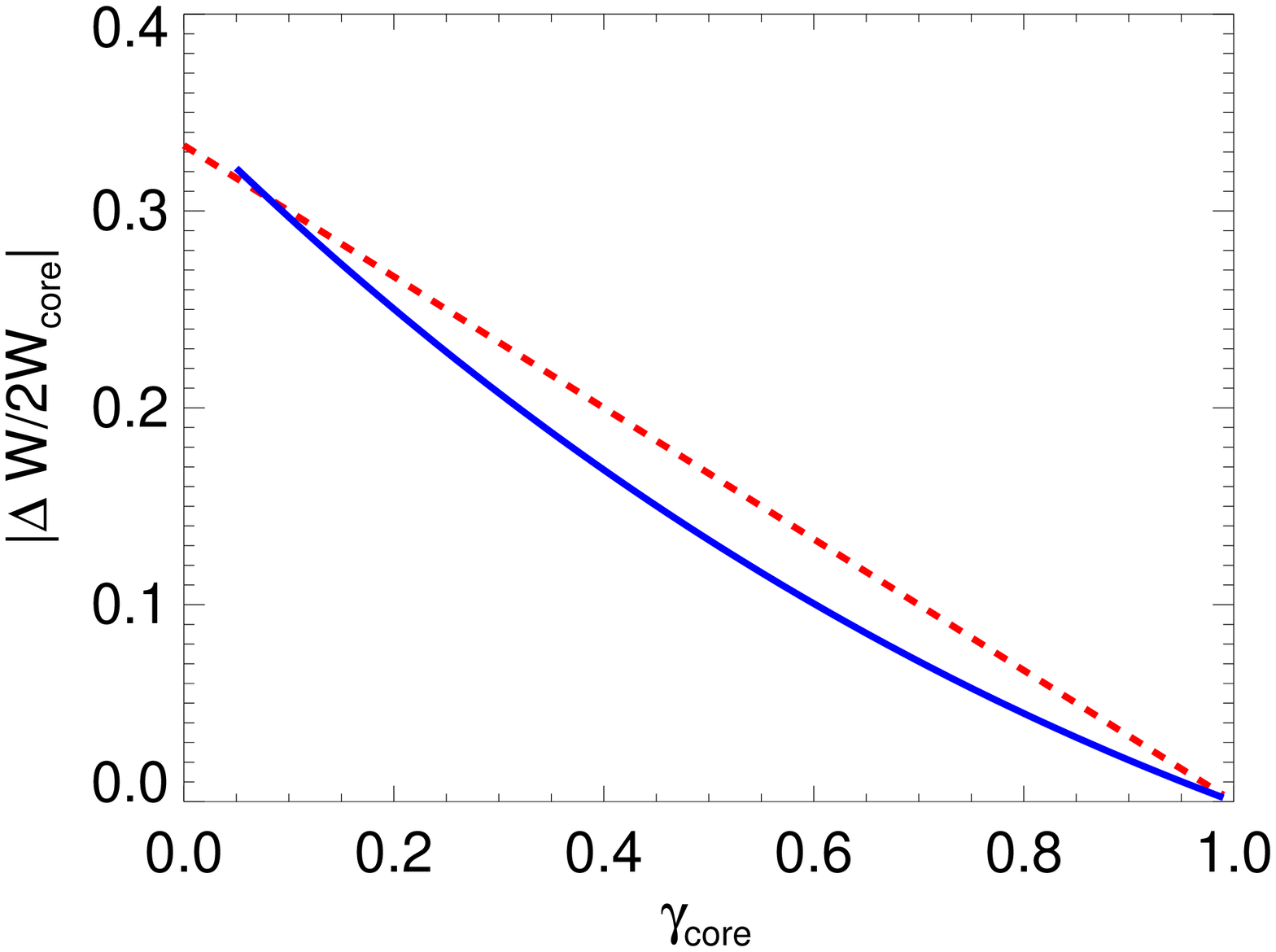}
\caption{Left: The prefactor $\zeta$ in Equation (\ref{zeta}) for DARKexp
(blue, solid) and the $\gamma$-models (red, dashed). Right: Relative energy 
required to transform an NFW cusp (logarithmic slope $-1$) into a core with 
central observed logarithmic slope of $\gamma_{\rm core}$. We plot 
$|\Delta W/2 W_{\rm core}|$ for DARKexp (blue, solid) and the $\gamma$-models
(red, dashed), given by Equations~(\ref{WDARK}) and (\ref{Wgamma}),
respectively. In both panels, $\gamma$ for DARKexp was obtained from
Equation~(\ref{gamma-approx}).}
\label{w}
\end{figure}

Cusps or cores of dark matter halos are governed by the normalized
dimensionless depth of the central potential well, to the extent that their 
potentials are shaped by collective relaxation. How different are halos with 
different normalized central potentials, apart from their central logarithmic 
density slopes? We obtain analytical expressions for their total potential 
energies in Appendix A. We can also express the total potential energy of a 
halo as
\begin{equation}
\label{zeta}
W=-\zeta \frac{GM^2}{r_h}
\end{equation}
with $\zeta\approx 0.4$ \citep{1969ApJ...158L.139S} (see Figure~\ref{w}). Here 
$M$ is the total mass (which is finite for DARKexp and the $\gamma$-models) 
and $r_h$ is the half-mass radius. 

For a $\gamma$-model, the halo binding energy relative to an NFW cusp model, 
i.e., $\gamma_{\rm cusp}=1$, is 
$| \Delta W/2 W_{\rm core} | = 1/3(1-\gamma_{\rm core})$ 
(Equation~\ref{Wgamma}). This relation is plotted in Figure~\ref{w} along with 
the corresponding relation based on DARKexp. The binding energy of a cored halo 
($\gamma_{\rm core}=0$) is $\sim 30\%$ less than that of a cuspy halo 
($\gamma_{\rm cusp}=1$), relative to the binding energy of the cored halo. For 
$\gamma_{\rm core}=0.5$, $\gamma_{\rm cusp}=1.5$ the difference is $\sim 25\%$. 
These differences are significant, i.e., halos do not have universal structures 
or relative binding energies if they have different central cusp strengths.

For a dwarf galaxy with virial mass $M=3\times10^9$ M$_\odot$, $r_h=10$ kpc, 
the total potential energy is $W\approx3\times10^{55}$ erg. The 
cusp--core transition therefore requires $\sim 10^{55}$ erg or $\sim 10^{4}$ 
supernovae (for a supernova energy input of $\sim 10^{51}$ erg) for 
$\gamma_{\rm core}=0$ (higher values of $\gamma_{\rm core}$ require less 
energy input). This corresponds to a total stellar mass produced of 
$\sim 10^{6}$ M$_\odot$, with an uncertainty of about a factor of 2 depending 
on the initial mass function. These estimates are consistent with those of 
\citet{2012ApJ...759L..42P} and \citet{2013MNRAS.433.3539G}. Because 
supernovae do not transfer all their kinetic energy into the dark matter, the 
energy injection is less efficient and more supernovae are needed. On the 
other hand, the energy input needed is diminished when considering the actual 
star formation and mass assembly history of dwarf spheroidals 
\citep{2014ApJ...782L..39A,2014ApJ...789L..17M}. Finally, observations do not 
strictly require $\gamma=0$, further reducing the energy requirements. 

As discussed in Section 3.2, dark matter halos of clusters of galaxies also 
appear to have central logarithmic slopes that are more shallow than $-1$
\citep[][]{2004ApJ...604...88S,2009ApJ...703L.132Z,2013ApJ...765...24N,2013ApJ...765...25N}.
The effect seems to be generic \citep[][]{2013ApJ...765...25N}, giving central 
logarithmic slopes in the range $0.2$--0.8. While some of this may be due to 
substructure, line of sight effects, or elongated structures (e.g., due to 
merging), it does appear that central profiles more shallow than NFW are 
observed in clusters of galaxies. However, for a massive cluster of galaxies 
with $M=1\times10^{15}$ M$_\odot$ and $r_h=500$ kpc, the energy input required 
is $\sim 10^{64}$ erg, corresponding to an unrealistic total stellar mass of 
$10^{15}$ M$_\odot$, if supernovae were to provide the energy input. It is 
therefore unlikely that supernova feedback is the origin of cores in clusters 
of galaxies. It is possible that dynamical friction \citep{2001ApJ...560..636E} 
of infalling satellites or black holes, active galactic nuclei 
\citep{2012MNRAS.422.3081M} or cluster mergers \citep{2013ApJ...765...25N} 
\citep[although see][]{2005MNRAS.360..892D} may provide the energy injection 
needed to erase the central cusp. Some clusters may also be born with a 
shallow potential, although it is noteworthy that a fair fraction of simulated 
cluster dark halos have central logarithmic slopes steeper than $-1$.

\section{Conclusions\label{conc}}

It has become standard practice to fit spherically averaged density profiles 
of dark matter halos, galaxies and clusters with phenomenological models, such 
as the NFW or its variants, or the Einasto $\alpha$-model. We suggest that the 
reason these functions fit well is because they happen to resemble DARKexp, 
which is a theoretically derived model for collisionless self-gravitating 
systems, based on statistical mechanical arguments.

Moreover, in this paper we have argued that non-universality of dark-matter 
halos is a natural expectation of DARKexp. If dark-matter halos would have 
logarithmic slopes at small radii robustly resolved in simulations equal 
to $-1$, this would require some fine tuning of the model. The expectation of 
a range of properties appears to be borne out in simulations, and in 
observations of dwarf galaxies, and clusters of galaxies. In this picture, the 
range of central logarithmic slopes inferred can be seen as a reflection 
of a range in their central normalized potentials, or, equivalently, their 
total normalized binding energies.  This insight is useful when quantifying 
the extent to which halos differ as a result of their formation history 
and also can be used to analytically relate halos with different cusps to each 
other, such as required in the cusp-core transition. We do not suggest that the 
central potential is set up by some external process which then in turn 
causally sets the density profile. Rather, the central potential and the 
logarithmic slope of the density profiles are both signatures of an 
underlying formation process which determines the final equilibrium state. 
Future data on both dwarfs and clusters will be necessary to more fully 
examine the applicability of DARKexp to the whole range of systems.

%%%
Because DARKexp does not have an analytical density profile, and if using 
tables of DARKexp $\rho(r)$ is not an option, we have identified 
phenomenological fitting functions that match DARKexp reasonably well, and have 
presented relations between DARKexp's $\phi_0$ and the best fitting $\gamma$ 
of the $\gamma$-model, and $\alpha$ of the Einasto model. The resulting DARKexp 
\citep{2010ApJ...722..851H} density profiles are well represented by 
Dehnen--Tremaine $\gamma$-models 
\citep{1993MNRAS.265..250D,1994AJ....107..634T}. The main advantage of these 
models is that they allow for a range of values of the central cusp logarithmic 
slope and so can be used to model both the standard NFW-type halos with 
$d\log \rho(r)/d\log r \approx -1$ as well as more shallow (or steep) halos. 
%%%

As shown in this paper (Figures~\ref{phi0-gamma} and \ref{density-sim}) 
and in other recent work 
\citep{2008MNRAS.387..536G,2008MNRAS.391.1685S,2013MNRAS.428.1696V,2013MNRAS.432.1103L}, 
cold dark matter numerical simulations by themselves generate a significant 
dispersion in $\alpha$ of simulated halos, reflecting different formation,
accretion, or tidal stripping histories. DARKexp naturally accounts for such 
non-universality in profile shapes. In Figure \ref{density-sim} we demonstrate 
the non-universality of the pure dark matter halos. Both the shallower and the 
steeper systems are well fit by DARKexp. While we have demonstrated a 
significant non-universality found in simulations of massive dark-matter halos, 
with a similar dispersion noticable in simulations of lower mass halos 
\citep[e.g.,][]{2013MNRAS.432.1103L}, it would be important to quantify the 
degree of non-universality and range of central potentials realized as a 
function of halo mass \citep[see also][]{2014MNRAS.441.2986D}.

%In Section~\ref{dwarfs} we compare the potential energies associated with 
In Section~4 we compare the potential energies associated with 
the core and cusp systems and discuss the required amount of energy and the 
possible injection mechanisms to bring about the transformation from a steeper 
pure dark matter system to a shallower observed system. The cusp--core 
transition cannot be simply modeled as a `one-size-fits-all' transition from 
an NFW halo (or a DARKexp with $\phi_0\sim 3-4$) to a cored halo (or a DARKexp 
with $\phi_0<2$), because for both dwarf galaxies \citep{2014ApJ...783....7C} 
and clusters \citep{2013ApJ...765...25N} there is a real dispersion in their 
observed properties. As highlighted in this paper the same is true for 
simulated massive dark-matter halos
\citep[see][for other mass ranges]{2013MNRAS.432.1103L, 2013MNRAS.431.1220D}. 
When modeling any transition of the central logarithmic slope properties 
as a result of some energy injection process, it is important to consider the 
significant dispersion in both the initial dark matter halos and those 
observed or affected by energy injection processes.

Finally, we note that the $\phi_0$--$\gamma$ relation (Figure~\ref{phi0-gamma}) 
can be directly tested in numerical simulations, e.g., dissipationless 
collapse or cosmological simulations. Halos with steeper profiles should have 
deeper central normalized potentials (possibly due to different formation times 
and accretion histories). For clusters of galaxies the $\phi_0$--$\gamma$ 
relation can be probed using gravitational redshift measurements 
\citep{2011Natur.477..567W} of the central potential and gravitational lensing 
measurements of the density profiles 
\citep{2011ApJ...738...41U,2013MNRAS.436.2616B,2015arXiv150704385U}.

\acknowledgments
We would like to thank Drew Newman for kindly providing us with their galaxy 
cluster fits and Nicola Amorisco, Arianna Di Cintio, Marceau Limousin, 
Keiichi Umetsu, and Jes\'us Zavala for comments. The Dark Cosmology Centre 
(DARK) is funded by the Danish National Research Foundation. L.L.R.W. would 
like to thank DARK for their hospitality. 

\appendix

\section{A. Potential energy of cored and cuspy DARK{\lowercase {exp}} and 
$\gamma$-models}

For DARKexp, $N(\epsilon)  = \exp(\phi_0-\epsilon)-1$,
\begin{equation}
M(\phi_0)=\int_0^{\phi_0} N(\epsilon) d\epsilon = \exp{\phi_0}-(\phi_0+1)
\end{equation}
and
\begin{equation}
K+2W=-\int_0^{\phi_0} \epsilon N(\epsilon) d\epsilon ,
\end{equation}
where $K$ is the total kinetic energy and $W$ is the total potential energy. 
Using the virial theorem, $W=-2K$,
\begin{equation}
\frac{3}{2} W=-M(\phi_0)+\frac{\phi_0^2}{2}.
\end{equation}
Assuming unit-mass DARKexp models, the energy difference between a cusp and 
a core relative to the current core is 
\begin{equation}\label{WDARK}
\left | \frac{\Delta W}{2 W_{\rm core}} \right | = \frac{1}{2} \left ( \frac{2-\left. \frac{\phi_0^2}{M(\phi_0)}\right |_{\rm cusp}}{2-\left. \frac{\phi_0^2}{M(\phi_0)}\right |_{\rm core}} -1 \right ).
\end{equation}

For a $\gamma$-model normalized to unit mass,
$\rho = \frac{3-\gamma}{4\pi}r^{-\gamma}(1+r)^{\gamma-4}$,
\begin{equation}
W=-(10-4\gamma)^{-1}
\end{equation}
and
\begin{equation}\label{Wgamma}
\left | \frac{\Delta W}{2 W_{\rm core}} \right | = 
\frac{1}{2} 
\left ( \frac{5-2\gamma_{\rm core}}{5-2\gamma_{\rm cusp}} -1 \right ).
\end{equation}

For the unit-mass $\gamma$-models, the relation between central potential and
inner logarithmic slope is 
\begin{equation}
\phi_\gamma=(2-\gamma)^{-1}.
\end{equation}
The relevant radii are also directly related to $\gamma$ through
\begin{equation}
r_{-2}=1-\gamma/2
\end{equation}
and the half-mass radius,
\begin{equation}
r_h= (2^{1/(3-\gamma)}-1)^{-1}
\end{equation}
\citep{1993MNRAS.265..250D}. Note that $\phi_\gamma r_{-2} = 0.5$ for 
$\gamma$-models; 
$\phi_0 r_{-2} \approx 0.5$ also holds for a unit mass DARKexp, 
further demonstrating that the two models are similar.

%%%
\section{B. Phenomenological fits to DARK{\lowercase {exp}}}

\begin{figure}%[t]
\epsscale{0.45}
\plotone{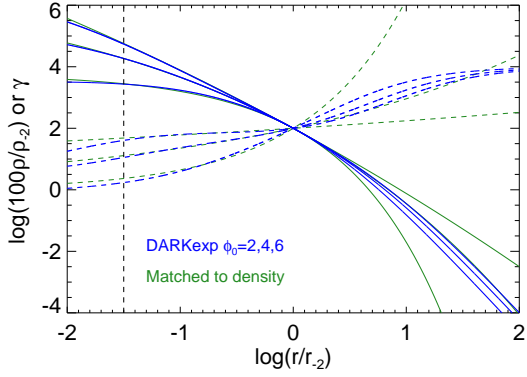}
\caption{
%%%
Comparison of DARKexp and Einasto models. DARKexp models with $\phi_0=2,4,6$ 
(from bottom to top at small radii) are shown in blue (solid curves: density 
profile; dashed curves: logarithmic density slope). In green we show 
$\alpha$-models matched at $\log (r_a/r_{-2}) = -1.5$ (dashed black lines) in 
density.
%%%
}
\label{einasto}
\end{figure}

Figure~\ref{einasto} shows a comparison between DARKexp and Einasto profiles. 
The correspondence is very good at small radii but poor at large radii. The 
poorer fit at larger radii is because the Einasto profile does not have a 
`built-in' outer logarithmic slope of $-4$, except for values around the 
canonical $\alpha=0.17$. The profiles were matched in density, so the 
corresponding blue and green curves intersect each other at $r_a$. The dashed 
curves in the same plot are the corresponding logarithmic density slopes; 
here too Einasto models track the respective DARKexp very well, although the 
blue and green dashed curves do not intersect each other at $r_a$ because the 
models can be matched either in density, or logarithmic density slope, 
not both.

\begin{figure}%[t]
\epsscale{0.45}
\plotone{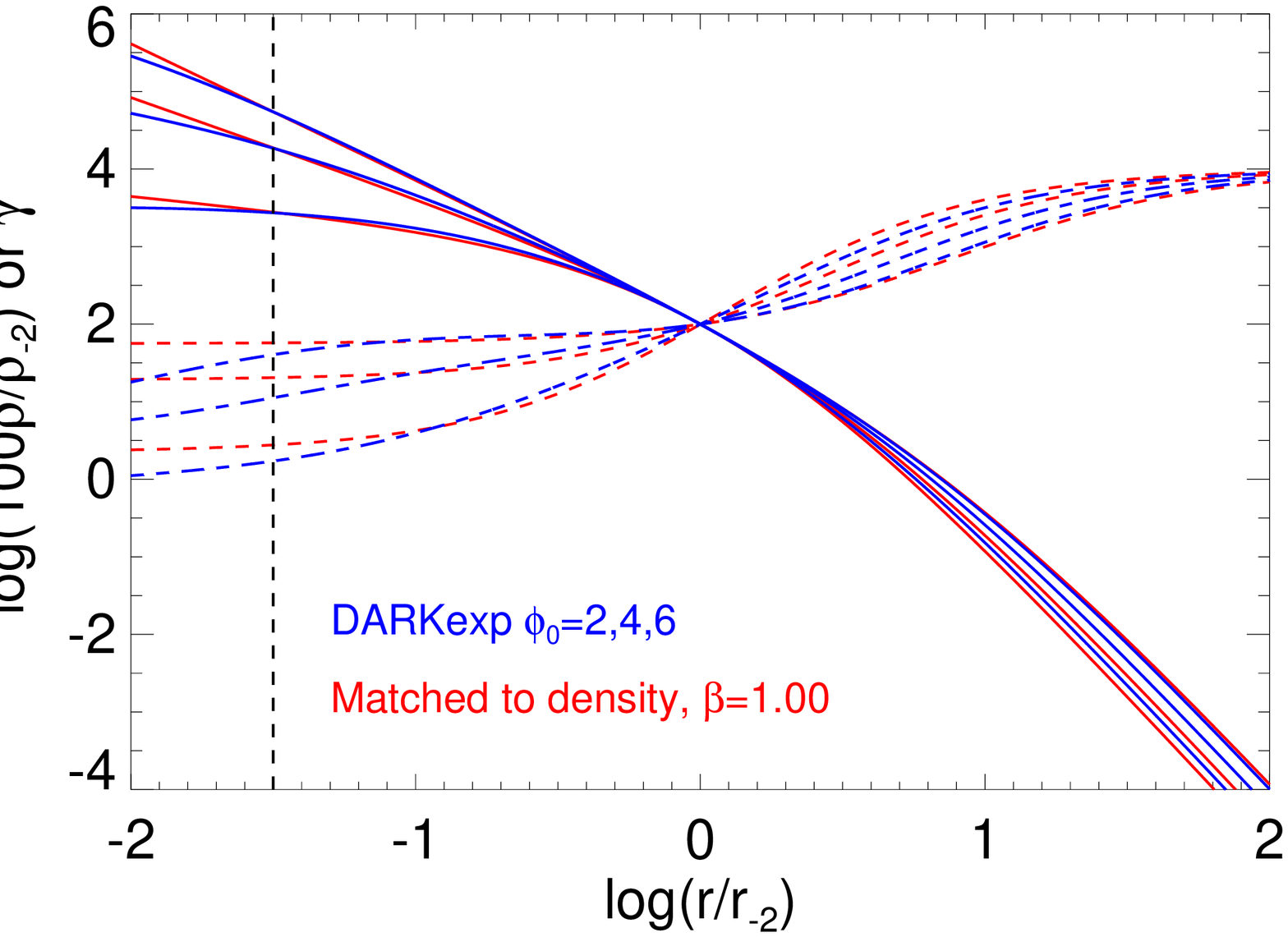}
\plotone{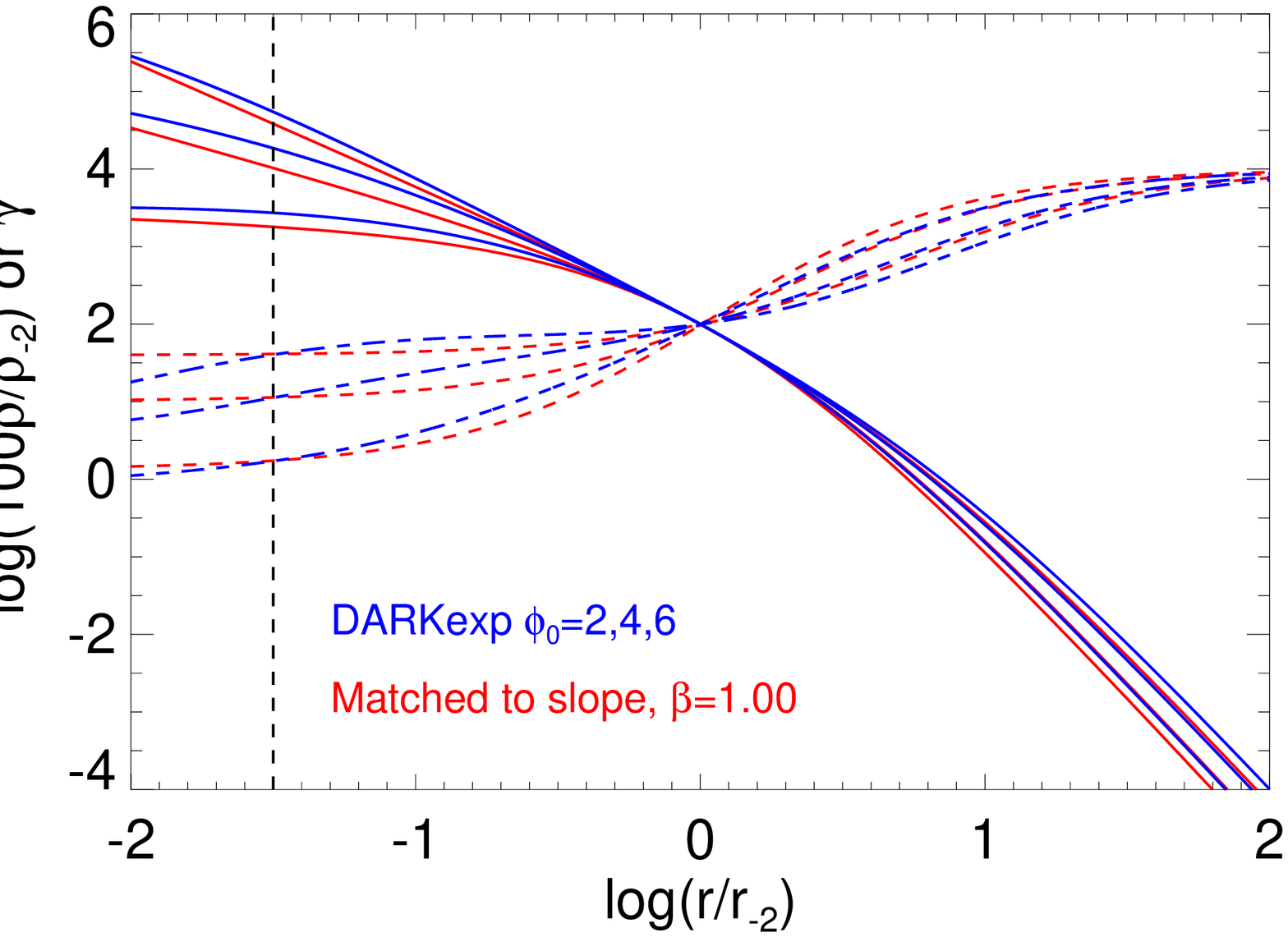}
\caption{
%%%
Comparison of DARKexp and $\gamma$-models (similar to Figure~2). In red we 
show $\gamma$-models (Hernquist models with $\beta=1$) matched at 
$\log (r_a/r_{-2}) = -1.5$ lines in either density (left) or logarithmic slope 
(right).
%%%
}
\label{benchmark}
\end{figure}

The left panel of Figure~\ref{benchmark} shows DARKexp of $\phi_0=2,4,6$ as 
solid blue curves, and the matched $\gamma$-models (Equation~\ref{gamod}) as 
red solid curves. These were matched in density. The $\gamma$-models 
approximate DARKexp very well over nearly 4 decades in radius. The right panel 
of Figure~\ref{benchmark} shows the same DARKexp density profiles with the 
$\gamma$-models that were matched in logarithmic density slope, at $r_a$. 
These too provide very good fits to DARKexp.

\begin{figure}%[t]
\epsscale{0.45}
\plotone{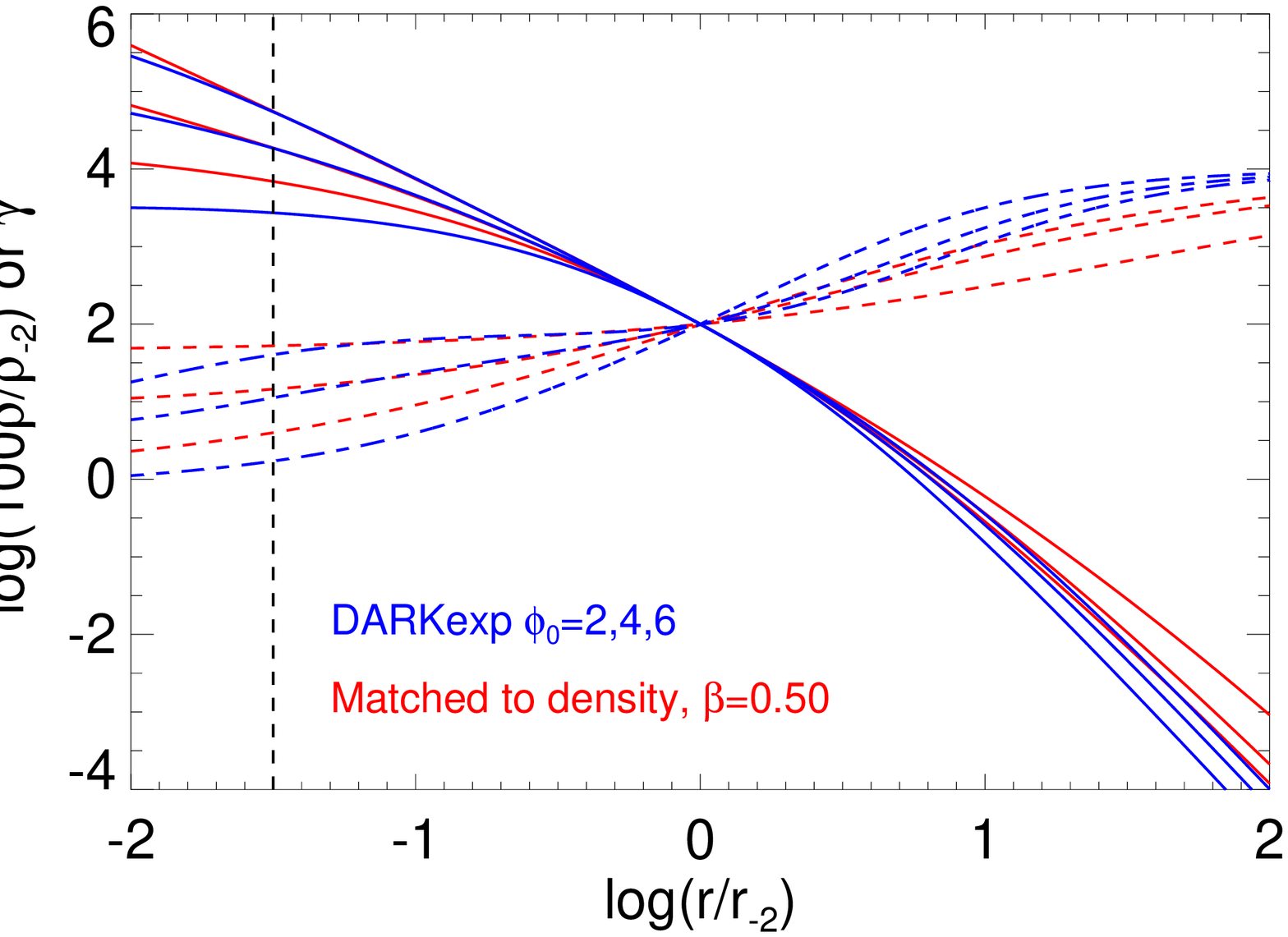}
\plotone{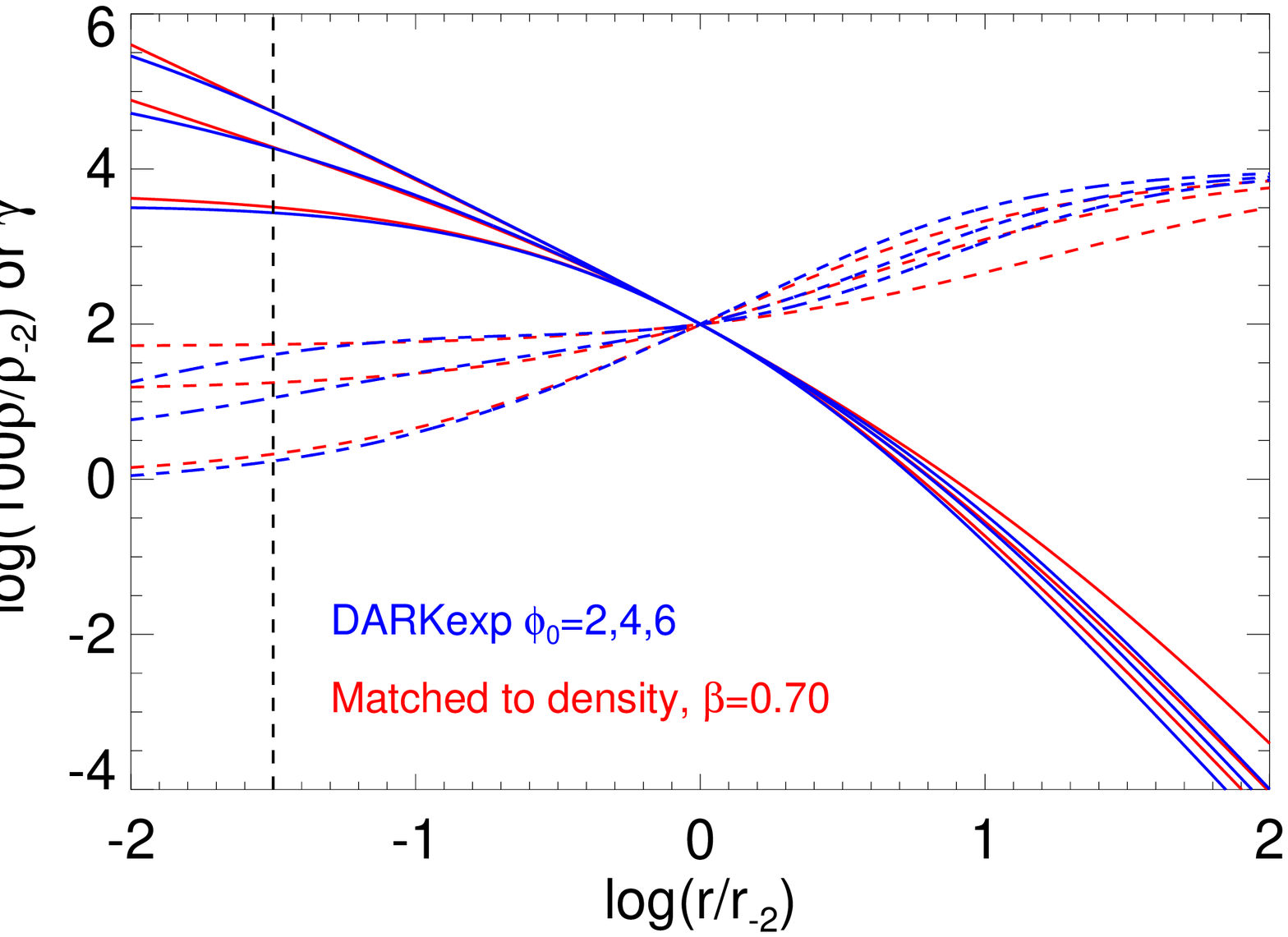}
\plotone{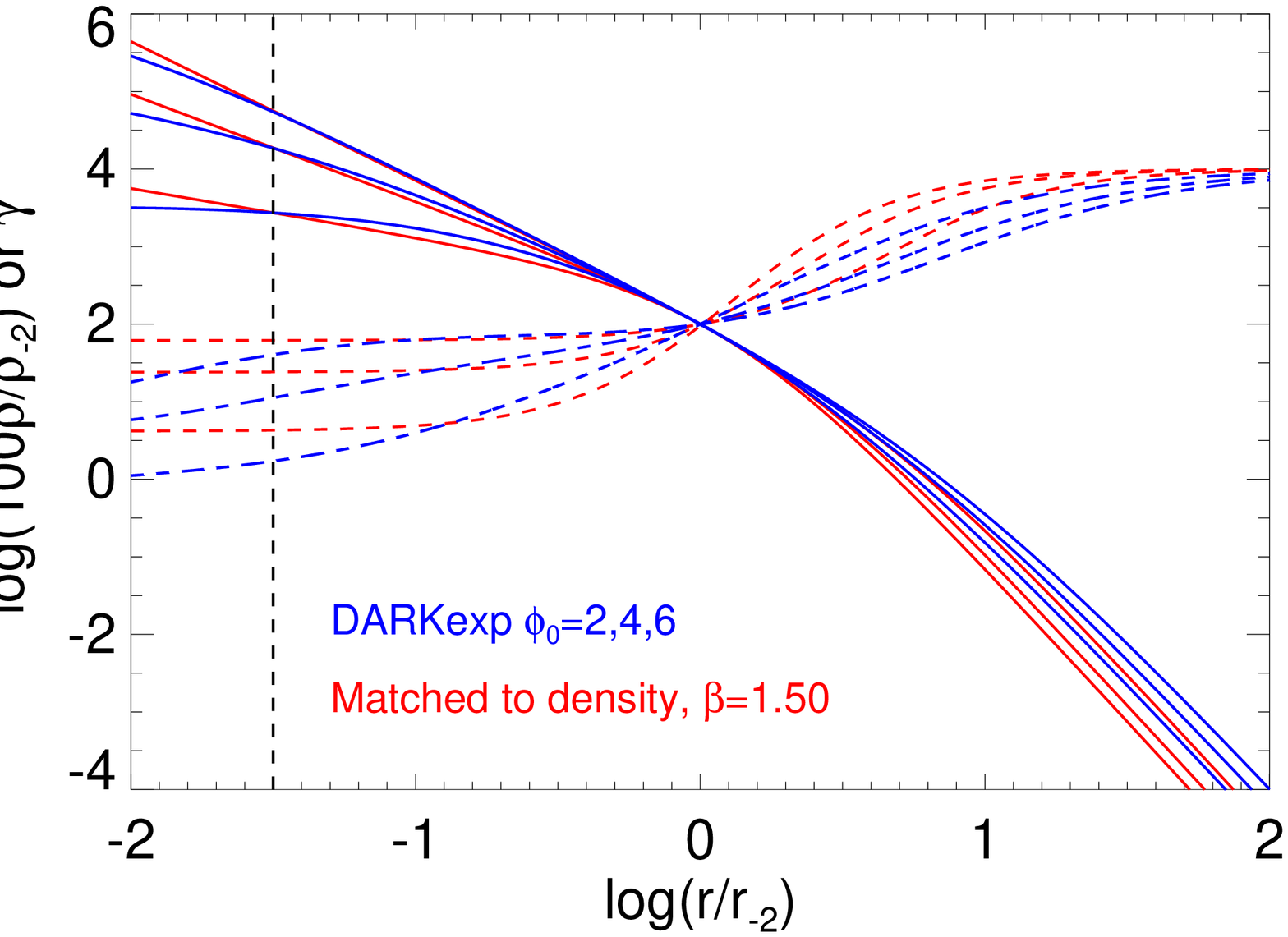}
\plotone{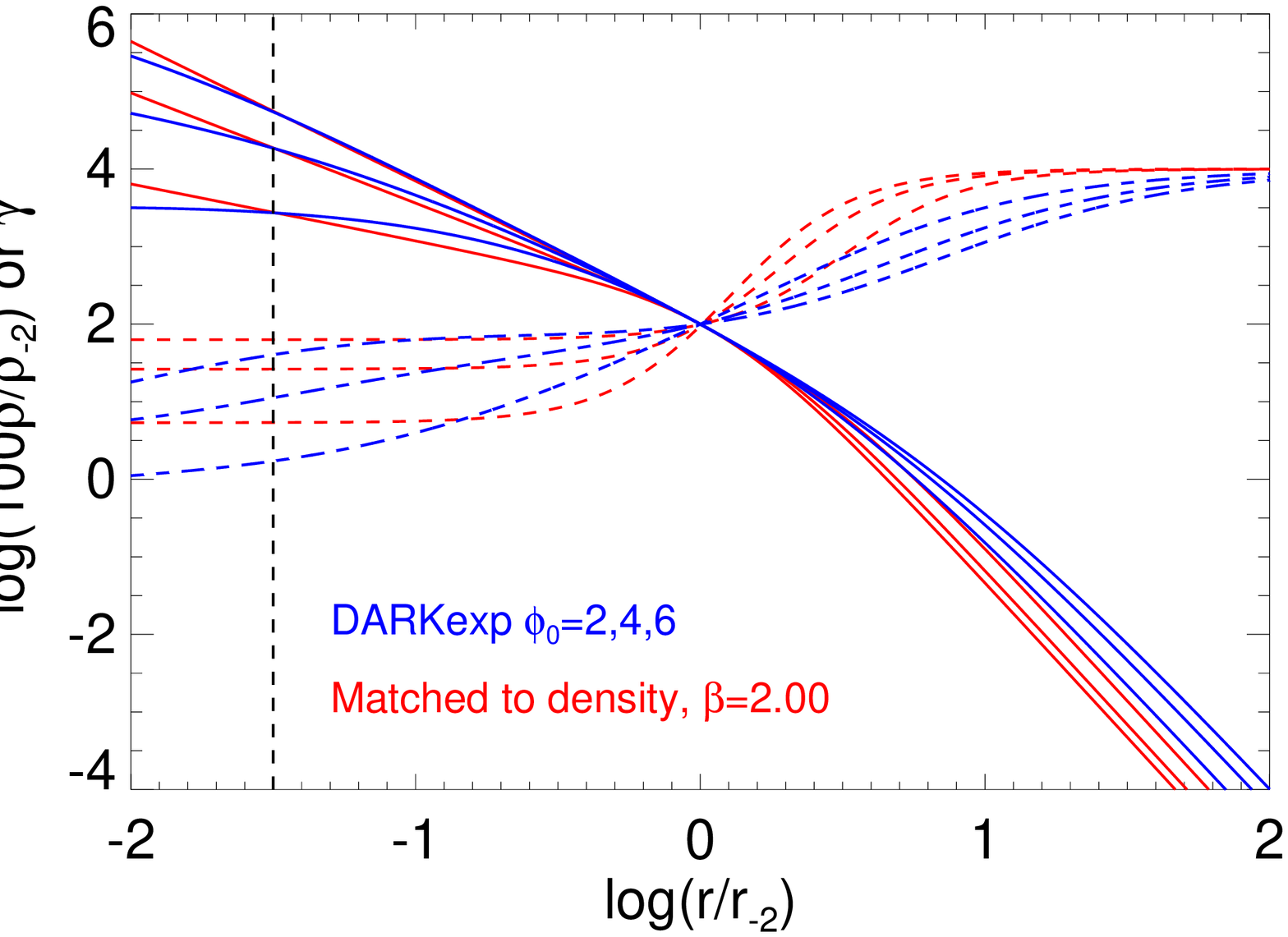}
\caption{
%%%
Effects of varying $\beta$ between 0.5 and 2.0. Matching is done in density 
at $\log (r/r_{-2}) = -1.5$ like in Figure~\ref{benchmark} (left).
%%%
}
\label{beta}
\end{figure}

Figure~\ref{beta} shows the effect of varying $\beta$ in the generalized 
Hernquist model (Equation~\ref{hernmod}), between $0.5$ and $2.0$ (here we 
show model pairs matched only in density, not logarithmic slope). A value of 
$\beta=0.7$ appears to provide a slightly better fit than $\beta=1$ but the 
improvement is not sufficiently significant to justify the introduction of 
another parameter. Note, however, that in the inner parts the correspondence 
with DARKexp is inferior to that of the Einasto profile, despite the fact that 
the generalized Hernquist model has an extra shape parameter. Overall, 
$\gamma$-models appear to be useful representations of DARKexp. 
%%%

\clearpage
\newpage
%\bibliography{phigamma}

\begin{thebibliography}{80}
\expandafter\ifx\csname natexlab\endcsname\relax\def\natexlab#1{#1}\fi

\bibitem[{{Adams} {et~al.}(2014){Adams}, {Simon}, {Fabricius}, {van den Bosch},
  {Barentine}, {Bender}, {Gebhardt}, {Hill}, {Murphy}, {Swaters}, {Thomas}, \&
  {van de Ven}}]{2014ApJ...789...63A}
{Adams}, J.~J., {et~al.} 2014, \apj, 789, 63

\bibitem[{{Amorisco} {et~al.}(2013){Amorisco}, {Agnello}, \&
  {Evans}}]{2013MNRAS.429L..89A}
{Amorisco}, N.~C., {Agnello}, A., \& {Evans}, N.~W. 2013, \mnras, 429, L89

\bibitem[{{Amorisco} \& {Evans}(2012)}]{2012MNRAS.419..184A}
{Amorisco}, N.~C., \& {Evans}, N.~W. 2012, \mnras, 419, 184

\bibitem[{{Amorisco} {et~al.}(2014){Amorisco}, {Zavala}, \& {de
  Boer}}]{2014ApJ...782L..39A}
{Amorisco}, N.~C., {Zavala}, J., \& {de Boer}, T.~J.~L. 2014, \apjl, 782, L39

\bibitem[{{Beraldo e Silva} {et~al.}(2013){Beraldo e Silva}, {Lima}, \&
  {Sodr{\'e}}}]{2013MNRAS.436.2616B}
{Beraldo e Silva}, L.~J., {Lima}, M., \& {Sodr{\'e}}, L. 2013, \mnras, 436,
  2616

\bibitem[{{Boylan-Kolchin} {et~al.}(2011){Boylan-Kolchin}, {Bullock}, \&
  {Kaplinghat}}]{2011MNRAS.415L..40B}
{Boylan-Kolchin}, M., {Bullock}, J.~S., \& {Kaplinghat}, M. 2011, \mnras, 415,
  L40

\bibitem[{{Brook} {et~al.}(2012){Brook}, {Stinson}, {Gibson}, {Wadsley}, \&
  {Quinn}}]{2012MNRAS.424.1275B}
{Brook}, C.~B., {Stinson}, G., {Gibson}, B.~K., {Wadsley}, J., \& {Quinn}, T.
  2012, \mnras, 424, 1275

\bibitem[{{Collins} {et~al.}(2014){Collins}, {Chapman}, {Rich}, {Ibata},
  {Martin}, {Irwin}, {Bate}, {Lewis}, {Pe{\~n}arrubia}, {Arimoto}, {Casey},
  {Ferguson}, {Koch}, {McConnachie}, \& {Tanvir}}]{2014ApJ...783....7C}
{Collins}, M.~L.~M., {et~al.} 2014, \apj, 783, 7

\bibitem[{{Dehnen}(1993)}]{1993MNRAS.265..250D}
{Dehnen}, W. 1993, \mnras, 265, 250

\bibitem[{{Dehnen}(2005)}]{2005MNRAS.360..892D}
---. 2005, \mnras, 360, 892

\bibitem[{{Di Cintio} {et~al.}(2014){Di Cintio}, {Brook}, {Dutton},
  {Macci{\`o}}, {Stinson}, \& {Knebe}}]{2014MNRAS.441.2986D}
{Di Cintio}, A., {Brook}, C.~B., {Dutton}, A.~A., {Macci{\`o}}, A.~V.,
  {Stinson}, G.~S., \& {Knebe}, A. 2014, \mnras, 441, 2986

\bibitem[{{Di Cintio} {et~al.}(2013){Di Cintio}, {Knebe}, {Libeskind}, {Brook},
  {Yepes}, {Gottl{\"o}ber}, \& {Hoffman}}]{2013MNRAS.431.1220D}
{Di Cintio}, A., {Knebe}, A., {Libeskind}, N.~I., {Brook}, C., {Yepes}, G.,
  {Gottl{\"o}ber}, S., \& {Hoffman}, Y. 2013, \mnras, 431, 1220

\bibitem[{{Diemand} \& {Moore}(2011)}]{2011ASL.....4..297D}
{Diemand}, J., \& {Moore}, B. 2011, Advanced Science Letters, 4, 297

\bibitem[{{Diemand} {et~al.}(2004){Diemand}, {Moore}, \&
  {Stadel}}]{2004MNRAS.353..624D}
{Diemand}, J., {Moore}, B., \& {Stadel}, J. 2004, \mnras, 353, 624

\bibitem[{{Diemand} {et~al.}(2005{\natexlab{a}}){Diemand}, {Moore}, \&
  {Stadel}}]{2005Natur.433..389D}
---. 2005{\natexlab{a}}, \nat, 433, 389

\bibitem[{{Diemand} {et~al.}(2005{\natexlab{b}}){Diemand}, {Zemp}, {Moore},
  {Stadel}, \& {Carollo}}]{2005MNRAS.364..665D}
{Diemand}, J., {Zemp}, M., {Moore}, B., {Stadel}, J., \& {Carollo}, C.~M.
  2005{\natexlab{b}}, \mnras, 364, 665

\bibitem[{{Einasto}(1965)}]{1965TrAlm...5...87E}
{Einasto}, J. 1965, Trudy Astrofizicheskogo Instituta Alma-Ata, 5, 87

\bibitem[{{El-Zant} {et~al.}(2001){El-Zant}, {Shlosman}, \&
  {Hoffman}}]{2001ApJ...560..636E}
{El-Zant}, A., {Shlosman}, I., \& {Hoffman}, Y. 2001, \apj, 560, 636

\bibitem[{{Flores} \& {Primack}(1994)}]{1994ApJ...427L...1F}
{Flores}, R.~A., \& {Primack}, J.~R. 1994, \apjl, 427, L1

\bibitem[{{Gao} {et~al.}(2008){Gao}, {Navarro}, {Cole}, {Frenk}, {White},
  {Springel}, {Jenkins}, \& {Neto}}]{2008MNRAS.387..536G}
{Gao}, L., {Navarro}, J.~F., {Cole}, S., {Frenk}, C.~S., {White}, S.~D.~M.,
  {Springel}, V., {Jenkins}, A., \& {Neto}, A.~F. 2008, \mnras, 387, 536

\bibitem[{{Garrison-Kimmel} {et~al.}(2013){Garrison-Kimmel}, {Rocha},
  {Boylan-Kolchin}, {Bullock}, \& {Lally}}]{2013MNRAS.433.3539G}
{Garrison-Kimmel}, S., {Rocha}, M., {Boylan-Kolchin}, M., {Bullock}, J.~S., \&
  {Lally}, J. 2013, \mnras, 433, 3539

\bibitem[{{Gilmore} {et~al.}(2007){Gilmore}, {Wilkinson}, {Wyse}, {Kleyna},
  {Koch}, {Evans}, \& {Grebel}}]{2007ApJ...663..948G}
{Gilmore}, G., {Wilkinson}, M.~I., {Wyse}, R.~F.~G., {Kleyna}, J.~T., {Koch},
  A., {Evans}, N.~W., \& {Grebel}, E.~K. 2007, \apj, 663, 948

\bibitem[{{Governato} {et~al.}(2012){Governato}, {Zolotov}, {Pontzen},
  {Christensen}, {Oh}, {Brooks}, {Quinn}, {Shen}, \&
  {Wadsley}}]{2012MNRAS.422.1231G}
{Governato}, F., {et~al.} 2012, \mnras, 422, 1231

\bibitem[{{Hernquist}(1990)}]{1990ApJ...356..359H}
{Hernquist}, L. 1990, \apj, 356, 359

\bibitem[{{Hjorth} \& {Madsen}(1991)}]{1991MNRAS.253..703H}
{Hjorth}, J., \& {Madsen}, J. 1991, \mnras, 253, 703

\bibitem[{{Hjorth} \& {Williams}(2010)}]{2010ApJ...722..851H}
{Hjorth}, J., \& {Williams}, L.~L.~R. 2010, \apj, 722, 851

\bibitem[{{Jaffe}(1987)}]{1987IAUS..127..511J}
{Jaffe}, W. 1987, in IAU Symposium, Vol. 127, Structure and Dynamics of
  Elliptical Galaxies, ed. P.~T. {de Zeeuw}, 511

\bibitem[{{Jardel} \& {Gebhardt}(2013)}]{2013arXiv1309.2637J}
{Jardel}, J.~R., \& {Gebhardt}, K. 2013, \apjl, 775, L30

\bibitem[{{Klypin} {et~al.}(2014){Klypin}, {Yepes}, {Gottlober}, {Prada}, \&
  {Hess}}]{2014arXiv1411.4001K}
{Klypin}, A., {Yepes}, G., {Gottlober}, S., {Prada}, F., \& {Hess}, S. 2014,
  arXiv:1411.4001

\bibitem[{{Klypin} {et~al.}(2011){Klypin}, {Trujillo-Gomez}, \&
  {Primack}}]{2011ApJ...740..102K}
{Klypin}, A.~A., {Trujillo-Gomez}, S., \& {Primack}, J. 2011, \apj, 740, 102

\bibitem[{{Limousin} {et~al.}(2008){Limousin}, {Richard}, {Kneib}, {Brink},
  {Pell{\'o}}, {Jullo}, {Tu}, {Sommer-Larsen}, {Egami}, {Micha{\l}owski},
  {Cabanac}, \& {Stark}}]{2008A&A...489...23L}
{Limousin}, M., {et~al.} 2008, \aap, 489, 23

\bibitem[{{Lovell} {et~al.}(2014){Lovell}, {Frenk}, {Eke}, {Jenkins}, {Gao}, \&
  {Theuns}}]{2014MNRAS.439..300L}
{Lovell}, M.~R., {Frenk}, C.~S., {Eke}, V.~R., {Jenkins}, A., {Gao}, L., \&
  {Theuns}, T. 2014, \mnras, 439, 300

\bibitem[{{Lovell} {et~al.}(2012){Lovell}, {Eke}, {Frenk}, {Gao}, {Jenkins},
  {Theuns}, {Wang}, {White}, {Boyarsky}, \& {Ruchayskiy}}]{2012MNRAS.420.2318L}
{Lovell}, M.~R., {et~al.} 2012, \mnras, 420, 2318

\bibitem[{{Ludlow} {et~al.}(2011){Ludlow}, {Navarro}, {White},
  {Boylan-Kolchin}, {Springel}, {Jenkins}, \& {Frenk}}]{2011MNRAS.415.3895L}
{Ludlow}, A.~D., {Navarro}, J.~F., {White}, S.~D.~M., {Boylan-Kolchin}, M.,
  {Springel}, V., {Jenkins}, A., \& {Frenk}, C.~S. 2011, \mnras, 415, 3895

\bibitem[{{Ludlow} {et~al.}(2013){Ludlow}, {Navarro}, {Boylan-Kolchin}, {Bett},
  {Angulo}, {Li}, {White}, {Frenk}, \& {Springel}}]{2013MNRAS.432.1103L}
{Ludlow}, A.~D., {et~al.} 2013, \mnras, 432, 1103

\bibitem[{{Macci{\`o}} {et~al.}(2012){Macci{\`o}}, {Paduroiu}, {Anderhalden},
  {Schneider}, \& {Moore}}]{2012MNRAS.424.1105M}
{Macci{\`o}}, A.~V., {Paduroiu}, S., {Anderhalden}, D., {Schneider}, A., \&
  {Moore}, B. 2012, \mnras, 424, 1105

\bibitem[{{Macci{\`o}} {et~al.}(2013){Macci{\`o}}, {Paduroiu}, {Anderhalden},
  {Schneider}, \& {Moore}}]{2013MNRAS.428.3715M}
---. 2013, \mnras, 428, 3715

\bibitem[{{Madau} {et~al.}(2014){Madau}, {Shen}, \&
  {Governato}}]{2014ApJ...789L..17M}
{Madau}, P., {Shen}, S., \& {Governato}, F. 2014, \apjl, 789, L17

\bibitem[{{Martizzi} {et~al.}(2012){Martizzi}, {Teyssier}, {Moore}, \&
  {Wentz}}]{2012MNRAS.422.3081M}
{Martizzi}, D., {Teyssier}, R., {Moore}, B., \& {Wentz}, T. 2012, \mnras, 422,
  3081

\bibitem[{{Merritt} {et~al.}(2006){Merritt}, {Graham}, {Moore}, {Diemand}, \&
  {Terzi{\'c}}}]{2006AJ....132.2685M}
{Merritt}, D., {Graham}, A.~W., {Moore}, B., {Diemand}, J., \& {Terzi{\'c}}, B.
  2006, \aj, 132, 2685

\bibitem[{{Moore}(1994)}]{1994Natur.370..629M}
{Moore}, B. 1994, \nat, 370, 629

\bibitem[{{Navarro} {et~al.}(1996){Navarro}, {Eke}, \&
  {Frenk}}]{1996MNRAS.283L..72N}
{Navarro}, J.~F., {Eke}, V.~R., \& {Frenk}, C.~S. 1996, \mnras, 283, L72

\bibitem[{{Navarro} {et~al.}(1997){Navarro}, {Frenk}, \&
  {White}}]{1997ApJ...490..493N}
{Navarro}, J.~F., {Frenk}, C.~S., \& {White}, S.~D.~M. 1997, \apj, 490, 493

\bibitem[{{Navarro} {et~al.}(2004){Navarro}, {Hayashi}, {Power}, {Jenkins},
  {Frenk}, {White}, {Springel}, {Stadel}, \& {Quinn}}]{2004MNRAS.349.1039N}
{Navarro}, J.~F., {et~al.} 2004, \mnras, 349, 1039

\bibitem[{{Navarro} {et~al.}(2010){Navarro}, {Ludlow}, {Springel}, {Wang},
  {Vogelsberger}, {White}, {Jenkins}, {Frenk}, \&
  {Helmi}}]{2010MNRAS.402...21N}
---. 2010, \mnras, 402, 21

\bibitem[{{Neto} {et~al.}(2007){Neto}, {Gao}, {Bett}, {Cole}, {Navarro},
  {Frenk}, {White}, {Springel}, \& {Jenkins}}]{2007MNRAS.381.1450N}
{Neto}, A.~F., {et~al.} 2007, \mnras, 381, 1450

\bibitem[{{Newman} {et~al.}(2013{\natexlab{a}}){Newman}, {Treu}, {Ellis}, \&
  {Sand}}]{2013ApJ...765...25N}
{Newman}, A.~B., {Treu}, T., {Ellis}, R.~S., \& {Sand}, D.~J.
  2013{\natexlab{a}}, \apj, 765, 25

\bibitem[{{Newman} {et~al.}(2013{\natexlab{b}}){Newman}, {Treu}, {Ellis},
  {Sand}, {Nipoti}, {Richard}, \& {Jullo}}]{2013ApJ...765...24N}
{Newman}, A.~B., {Treu}, T., {Ellis}, R.~S., {Sand}, D.~J., {Nipoti}, C.,
  {Richard}, J., \& {Jullo}, E. 2013{\natexlab{b}}, \apj, 765, 24

\bibitem[{{O{\~n}orbe} {et~al.}(2015){O{\~n}orbe}, {Boylan-Kolchin}, {Bullock},
  {Hopkins}, {Ker{\v e}s}, {Faucher-Gigu{\`e}re}, {Quataert}, \&
  {Murray}}]{2015arXiv150202036O}
{O{\~n}orbe}, J., {Boylan-Kolchin}, M., {Bullock}, J.~S., {Hopkins}, P.~F.,
  {Ker{\v e}s}, D., {Faucher-Gigu{\`e}re}, C.-A., {Quataert}, E., \& {Murray},
  N. 2015, arXiv:1502.02036

\bibitem[{{Pe{\~n}arrubia} {et~al.}(2012){Pe{\~n}arrubia}, {Pontzen}, {Walker},
  \& {Koposov}}]{2012ApJ...759L..42P}
{Pe{\~n}arrubia}, J., {Pontzen}, A., {Walker}, M.~G., \& {Koposov}, S.~E. 2012,
  \apjl, 759, L42

\bibitem[{{Pontzen} \& {Governato}(2012)}]{2012MNRAS.421.3464P}
{Pontzen}, A., \& {Governato}, F. 2012, \mnras, 421, 3464

\bibitem[{{Pontzen} \& {Governato}(2014)}]{2014Natur.506..171P}
---. 2014, \nat, 506, 171

\bibitem[{{Power} {et~al.}(2003){Power}, {Navarro}, {Jenkins}, {Frenk},
  {White}, {Springel}, {Stadel}, \& {Quinn}}]{2003MNRAS.338...14P}
{Power}, C., {Navarro}, J.~F., {Jenkins}, A., {Frenk}, C.~S., {White},
  S.~D.~M., {Springel}, V., {Stadel}, J., \& {Quinn}, T. 2003, \mnras, 338, 14

\bibitem[{{Riebe} {et~al.}(2013){Riebe}, {Partl}, {Enke}, {Forero-Romero},
  {Gottl{\"o}ber}, {Klypin}, {Lemson}, {Prada}, {Primack}, {Steinmetz}, \&
  {Turchaninov}}]{2013AN....334..691R}
{Riebe}, K., {et~al.} 2013, Astronomische Nachrichten, 334, 691

\bibitem[{{Rocha} {et~al.}(2013){Rocha}, {Peter}, {Bullock}, {Kaplinghat},
  {Garrison-Kimmel}, {O{\~n}orbe}, \& {Moustakas}}]{2013MNRAS.430...81R}
{Rocha}, M., {Peter}, A.~H.~G., {Bullock}, J.~S., {Kaplinghat}, M.,
  {Garrison-Kimmel}, S., {O{\~n}orbe}, J., \& {Moustakas}, L.~A. 2013, \mnras,
  430, 81

\bibitem[{{Sand} {et~al.}(2004){Sand}, {Treu}, {Smith}, \&
  {Ellis}}]{2004ApJ...604...88S}
{Sand}, D.~J., {Treu}, T., {Smith}, G.~P., \& {Ellis}, R.~S. 2004, \apj, 604,
  88

\bibitem[{{Shao} {et~al.}(2013){Shao}, {Gao}, {Theuns}, \&
  {Frenk}}]{2013MNRAS.430.2346S}
{Shao}, S., {Gao}, L., {Theuns}, T., \& {Frenk}, C.~S. 2013, \mnras, 430, 2346

\bibitem[{{Shen} {et~al.}(2014){Shen}, {Madau}, {Conroy}, {Governato}, \&
  {Mayer}}]{2014ApJ...792...99S}
{Shen}, S., {Madau}, P., {Conroy}, C., {Governato}, F., \& {Mayer}, L. 2014,
  \apj, 792, 99

\bibitem[{{Spitzer}(1969)}]{1969ApJ...158L.139S}
{Spitzer}, Jr., L. 1969, \apjl, 158, L139

\bibitem[{{Springel} {et~al.}(2008){Springel}, {Wang}, {Vogelsberger},
  {Ludlow}, {Jenkins}, {Helmi}, {Navarro}, {Frenk}, \&
  {White}}]{2008MNRAS.391.1685S}
{Springel}, V., {et~al.} 2008, \mnras, 391, 1685

\bibitem[{{Strigari} {et~al.}(2006){Strigari}, {Bullock}, {Kaplinghat},
  {Kravtsov}, {Gnedin}, {Abazajian}, \& {Klypin}}]{2006ApJ...652..306S}
{Strigari}, L.~E., {Bullock}, J.~S., {Kaplinghat}, M., {Kravtsov}, A.~V.,
  {Gnedin}, O.~Y., {Abazajian}, K., \& {Klypin}, A.~A. 2006, \apj, 652, 306

\bibitem[{{Strigari} {et~al.}(2014){Strigari}, {Frenk}, \&
  {White}}]{2014arXiv1406.6079S}
{Strigari}, L.~E., {Frenk}, C.~S., \& {White}, S.~D.~M. 2014, arXiv:1406.6079

\bibitem[{{Tasitsiomi} {et~al.}(2004){Tasitsiomi}, {Kravtsov}, {Gottl{\"o}ber},
  \& {Klypin}}]{2004ApJ...607..125T}
{Tasitsiomi}, A., {Kravtsov}, A.~V., {Gottl{\"o}ber}, S., \& {Klypin}, A.~A.
  2004, \apj, 607, 125

\bibitem[{{Tremaine} {et~al.}(1994){Tremaine}, {Richstone}, {Byun}, {Dressler},
  {Faber}, {Grillmair}, {Kormendy}, \& {Lauer}}]{1994AJ....107..634T}
{Tremaine}, S., {Richstone}, D.~O., {Byun}, Y.-I., {Dressler}, A., {Faber},
  S.~M., {Grillmair}, C., {Kormendy}, J., \& {Lauer}, T.~R. 1994, \aj, 107, 634

\bibitem[{{Umetsu} {et~al.}(2011){Umetsu}, {Broadhurst}, {Zitrin},
  {Medezinski}, {Coe}, \& {Postman}}]{2011ApJ...738...41U}
{Umetsu}, K., {Broadhurst}, T., {Zitrin}, A., {Medezinski}, E., {Coe}, D., \&
  {Postman}, M. 2011, \apj, 738, 41

\bibitem[{{Umetsu} {et~al.}(2015){Umetsu}, {Zitrin}, {Gruen}, {Merten},
  {Donahue}, \& {Postman}}]{2015arXiv150704385U}
  {Umetsu}, K., {Zitrin}, A., {Gruen}, D., {Merten}, J., {Donahue}, M., \&
    {Postman}, M. 2015, arXiv:1507.04385

\bibitem[{{Vera-Ciro} {et~al.}(2013){Vera-Ciro}, {Helmi}, {Starkenburg}, \&
  {Breddels}}]{2013MNRAS.428.1696V}
{Vera-Ciro}, C.~A., {Helmi}, A., {Starkenburg}, E., \& {Breddels}, M.~A. 2013,
  \mnras, 428, 1696

\bibitem[{{Vogelsberger} {et~al.}(2012){Vogelsberger}, {Zavala}, \&
  {Loeb}}]{2012MNRAS.423.3740V}
{Vogelsberger}, M., {Zavala}, J., \& {Loeb}, A. 2012, \mnras, 423, 3740

\bibitem[{{Vogelsberger} {et~al.}(2014){Vogelsberger}, {Zavala}, {Simpson}, \&
  {Jenkins}}]{2014MNRAS.444.3684V}
{Vogelsberger}, M., {Zavala}, J., {Simpson}, C., \& {Jenkins}, A. 2014, \mnras,
  444, 3684

\bibitem[{{Walker} {et~al.}(2009){Walker}, {Mateo}, {Olszewski},
  {Pe{\~n}arrubia}, {Wyn Evans}, \& {Gilmore}}]{2009ApJ...704.1274W}
{Walker}, M.~G., {Mateo}, M., {Olszewski}, E.~W., {Pe{\~n}arrubia}, J., {Wyn
  Evans}, N., \& {Gilmore}, G. 2009, \apj, 704, 1274

\bibitem[{{Walker} \& {Pe{\~n}arrubia}(2011)}]{2011ApJ...742...20W}
{Walker}, M.~G., \& {Pe{\~n}arrubia}, J. 2011, \apj, 742, 20

\bibitem[{{Weinberg} {et~al.}(2013){Weinberg}, {Bullock}, {Governato}, {Kuzio
  de Naray}, \& {Peter}}]{2013arXiv1306.0913W}
{Weinberg}, D.~H., {Bullock}, J.~S., {Governato}, F., {Kuzio de Naray}, R., \&
  {Peter}, A.~H.~G. 2013, arXiv:1306.0913

\bibitem[{{Williams} {et~al.}(2012){Williams}, {Barnes}, \&
  {Hjorth}}]{2012MNRAS.423.3589W}
{Williams}, L.~L.~R., {Barnes}, E.~I., \& {Hjorth}, J. 2012, \mnras, 423, 3589

\bibitem[{{Williams} \& {Hjorth}(2010)}]{2010ApJ...722..856W}
{Williams}, L.~L.~R., \& {Hjorth}, J. 2010, \apj, 722, 856

\bibitem[{{Williams} {et~al.}(2010){Williams}, {Hjorth}, \&
  {Wojtak}}]{2010ApJ...725..282W}
{Williams}, L.~L.~R., {Hjorth}, J., \& {Wojtak}, R. 2010, \apj, 725, 282

\bibitem[{{Williams} {et~al.}(2014){Williams}, {Hjorth}, \&
  {Wojtak}}]{2014ApJ...783...13W}
---. 2014, \apj, 783, 13

\bibitem[{{Wojtak} {et~al.}(2011){Wojtak}, {Hansen}, \&
  {Hjorth}}]{2011Natur.477..567W}
{Wojtak}, R., {Hansen}, S.~H., \& {Hjorth}, J. 2011, \nat, 477, 567

\bibitem[{{Zavala} {et~al.}(2013){Zavala}, {Vogelsberger}, \&
  {Walker}}]{2013MNRAS.431L..20Z}
{Zavala}, J., {Vogelsberger}, M., \& {Walker}, M.~G. 2013, \mnras, 431, L20

\bibitem[{{Zitrin} \& {Broadhurst}(2009)}]{2009ApJ...703L.132Z}
{Zitrin}, A., \& {Broadhurst}, T. 2009, \apjl, 703, L132

\bibitem[{{Zolotov} {et~al.}(2012){Zolotov}, {Brooks}, {Willman}, {Governato},
  {Pontzen}, {Christensen}, {Dekel}, {Quinn}, {Shen}, \&
  {Wadsley}}]{2012ApJ...761...71Z}
{Zolotov}, A., {et~al.} 2012, \apj, 761, 71

\end{thebibliography}

\clearpage

\end{document}